\documentclass[aip,jcp,amsmath,amssymb,
preprint,groupedaddress]{revtex4-1}
\usepackage{graphicx}
\usepackage{dcolumn}
\usepackage{bbm}
\usepackage{subcaption}
\usepackage[utf8]{inputenc}
\usepackage[T1]{fontenc}
\usepackage{mathptmx}
\usepackage {CJK}
\usepackage{xcolor}

\begin{document}
\begin {CJK*}{UTF8}{gbsn} 
\title[]{Identification of a Multi-Dimensional Reaction Coordinate for Crystal Nucleation in $\text{Ni}_3\text{Al}$}
\author{Yanyan Liang(梁燕燕)}
\email{yanyan.liang@rub.de}
\author{Grisell D\'{i}az Leines}
\author{Ralf Drautz}
\author{Jutta Rogal}
\email{jutta.rogal@rub.de}
\affiliation{Interdisciplinary Centre for Advanced Materials Simulation, Ruhr-Universit{\"a}t Bochum, 44780 Bochum, Germany}
\date{\today}
\begin{abstract}
Nucleation during solidification in multi-component alloys is a complex process that comprises the competition between different crystalline phases as well as chemical composition and ordering.
Here, we combine transition interface sampling with an extensive committor analysis to investigate the atomistic mechanisms during the initial stages of nucleation in $\text{Ni}_3\text{Al}$. The formation and growth of crystalline clusters from the melt are strongly influenced by the interplay between three descriptors: the size, crystallinity, and chemical short-range order of the emerging nuclei. We demonstrate that it is essential to include all three features in a multi-dimensional reaction coordinate to correctly describe the nucleation mechanism, where in particular the chemical short-range order plays a crucial role in the stability of small clusters.
The necessity of identifying multi-dimensional reaction coordinates is expected to be of key importance for the atomistic characterization of nucleation processes in complex, multi-component systems.
\end{abstract}
\maketitle
\end{CJK*}


\section{Introduction}
\label{sec:intro}
Fundamental knowledge of crystal nucleation in multi-component systems is essential for the controlled synthesis of materials with targeted properties. Despite this, insight into the atomistic mechanisms of nucleation remains elusive as even simple model systems exhibit complex transitions that involve several steps, polymorphic structures, and multiple order parameters.~\cite{sosso2016crystal,russo2016crystal} For multi-component systems, the additional complexity that emerges from possible ordering tendencies of the chemical species poses a major challenge for both experiment and theory. Consequently, even less is known about the underlying mechanisms that govern the initial steps of crystallization in technological relevant alloys.

Characterizing the mechanism of a complex, activated process like nucleation is intrinsically linked to identifying the degrees of freedom that represent a relevant reaction coordinate (RC). On the mesoscale, classical nucleation theory (CNT)~\cite{becker1935kinetische,binder1987theory} has been successfully applied to study nucleation. One of the key assumptions in CNT is that crystal nucleation is a one-dimensional process described entirely by a single RC, namely the radius of a spherical nucleus or, more generally, the size of the growing cluster.
However, several theoretical studies have shown that even for simple systems like Lennard-Jones (LJ) liquids, hard spheres, and colloidal suspensions, a one-dimensional RC is not sufficient to correctly capture the nucleation mechanism.~\cite{pan2004dynamics,moroni2005interplay,trudu2006freezing,schilling2010precursor,beckham2011optimizing,lechner2011reaction,russo2016crystal,berendsen2019unbiased}. Indeed, crystal nucleation is inherently a multi-dimensional process which has been described as \emph{ the ordering of multiple order parameters}~\cite{russo2016crystal}, where a trade-off between entropy and enthalpy~\cite{piaggi2017enhancing} characterizes the transition, and a decoupling of translational and orientational order can occur at different temperatures~\cite{russo2016crystal}. In binary or multi-component systems where changes in chemical order and composition, as well as demixing can play a key role during nucleation and growth, the identification of meaningful, low-dimensional collective variables (CVs) that can serve as RCs is extremely challenging.

Theoretically, the committor~\cite{bolhuis2002transition,du1998transition,geissler1999kinetic} can be considered as the optimal RC of an activated process.~\cite{peters2006obtaining}  The committor provides a statistical measure of the progress of a transition between two states by measuring the probability for any given configuration to commit to the final state. Due to its statistical definition the committor does not yield direct physical insight into the mechanism, yet it can be used to evaluate the quality of physically meaningful collective variables as RCs. Good CV candidates for the reaction coordinate should, for example, exhibit a strong correlation with the committor.
For studies of crystal nucleation in single-component systems such as LJ systems, soft core colloids and pure metals~\cite{moroni2005interplay,lechner2011role,diaz2018maximum}, extensive committor analysis revealed that the size of the crystalline nucleus is not sufficient to model the RC, and additional variables such as the crystallinity or a cloud of pre-structured surface particles around a crystalline core improve the description of the mechanism.

Several theoretical studies have investigated crystal nucleation in two-component systems, including bimetallics of Pd-Ag, Cu-Ni, Pd-Ni, Cu-Pd~\cite{desgranges2014unraveling,desgranges2016effect,bechelli2017free,watson2011crystal}, and mixtures of colloidal particles~\cite{anwar1998computer,gruhn2001molecular,mucha2003salt,zahn2004atomistic,punnathanam2006crystal,sanz2007evidence,peters2009competing}. The bimetallic systems above have rather simple phase diagrams forming solid solution bulk phases without strong chemical ordering tendencies. Consequently, the main focus of these studies is on demixing and the enrichment of chemical species in the growing cluster~\cite{desgranges2014unraveling,desgranges2016effect,bechelli2017free,watson2011crystal}. The initial crystal nucleation process can, however, be considerably different for chemically ordered phases compared to random solid solutions.
Indeed, studies of crystal nucleation in colloidal mixtures showed that the nucleation mechanism is largely affected by the underlying complex phase diagram containing ordered compounds. Competing structures of CsCl ordered body-centered cubic (bcc) and disordered face-centered cubic (fcc) were found during nucleation~\cite{punnathanam2006crystal,sanz2007evidence,peters2009competing} and structure-specific parameters were required to differentiate competing pathways that lead to different bulk phases~\cite{peters2009competing}.

In this work we investigate the mechanisms of crystal nucleation in binary $\text{Ni}_3\text{Al}$. Ni-Al alloys are of particular interest since they serve as a basis for high-performance materials that are key in many technological applications~\cite{pope1996high}. 
In contrast to the bimetallic alloys mentioned previously, Ni-Al exhibits a fairly complicated phase diagram with a number of chemically ordered structures~\cite{xu1990phase,massalski1992binary,huang1998thermodynamic}. 
Specifically, Ni-Al exhibits a single phase region of $\text{Ni}_3\text{Al}$ ranging from 73 at.\% to 77 at.\% Ni in the equilibrium phase diagram~\cite{massalski1992binary} where it forms an $\text{L1}_2$ ordered fcc structure that is thermodynamically stable up to the melting temperature.
It was, however, shown in electromagnetic levitation experiments that during rapid solidification in Ni-rich alloys with 23.5-30.8 at.\% Al there is a strong competition between the formation of several chemically ordered and disordered fcc and bcc phases.~\cite{assadi1998kinetics}
But it remains unclear how this competition between various chemically ordered and disordered phases influences the nucleation mechanism, at what stage of the nucleation process this becomes relevant, and how it is reflected in a set of CVs needed to characterize the mechanism.
To address these open questions we have combined transition interface sampling (TIS)~\cite{van2005elaborating},
for an enhanced sampling of the nucleation process in $\text{Ni}_3\text{Al}$, with a committor analysis~\cite{dellago2002transition,bolhuis2002transition} to assess the quality of proposed CVs as reaction coordinates. We find that during crystal nucleation in $\text{Ni}_3\text{Al}$ there is indeed a competition between various crystalline structures, strongly indicating the existence of multiple reaction channels. As a consequence, the size of the growing cluster as single RC results in an incomplete description of the nucleation mechanism, in contrast to the nucleation in unary metals~\cite{diaz2017atomistic,desgranges2007molecular}.
In addition to the size, information concerning the crystallinity as well as the chemical short-range order (SRO) are required to differentiate potential nucleation pathways. In particular, the chemical SRO is discovered to be a crucial factor triggering continuous growth or shrinkage of solid clusters with the same size and crystallinity. The initial stage of crystal nucleation in $\text{Ni}_3\text{Al}$ exhibits an enormous complexity caused by the interplay between size, crystallinity, and short-range order. As a result, all of these aspects need to be included in the RC for an accurate representation of the nucleation process. We expect that generally in complex systems with competing nucleation pathways, the assessment of multi-dimensional reaction coordinates is of fundamental importance.

\section{Methods}
\label{sec:methods}
\subsection{Committor Analysis}
\label{subsec:CommtAnal}
An approach widely used to analyze reaction coordinates is based on the concept of the commitment probability or committor, $p_{B}$.~\cite{bolhuis2002transition,du1998transition,geissler1999kinetic} In a system with two metastable states, $A$ and $B$, the committor $p_B(\mathbf{r})$ is defined as the probability that a trajectory starting at a configuration $\mathbf{r}$ reaches state $B$ before $A$, calculated as an average over the Boltzmann distributed momenta at a given temperature.
The committor provides a statistical measure of the progress of the reaction from the initial state $A$ to the final state $B$ and is often considered the perfect RC.~\cite{peters2006obtaining} By definition, $p_B(\mathbf{r}) = 0$ for configurations within state $A$ and $p_B(\mathbf{r}) = 1$ for configurations within $B$.  Configurations with $p_B(\mathbf{r}) = 0.5$ mark the separatix or transition state (TS).
The committor does not, however, yield any direct physical insight into the mechanisms of reactive events, but it can be used to evaluate the quality of collective variables, $\mathbf{q}(\mathbf{r}) = \{q_1(\mathbf{r}),\dots,q_n(\mathbf{r})\}$, as reaction coordinates. 
Specifically, for a \emph{good} reaction coordinate the CVs should be strongly correlated with the committor.
The quality of any given vector $\mathbf{q}(\mathbf{r})$ of CV values as reaction coordinate can be assessed on the basis of the committor distribution~\cite{dellago2002transition,bolhuis2002transition}
\begin{equation}
 \label{eq:committor_dist}
P(p_B|_\mathbf{q}) = \frac{\langle \delta \left( p_B(\mathbf{r}) - p_B \right)  \delta\left( \mathbf{q}(\mathbf{r}) - \mathbf{q} \right) \rangle}{\langle \delta\left( \mathbf{q}(\mathbf{r}) - \mathbf{q} \right) \rangle} \quad .
\end{equation}
Here, $\langle \dots \rangle$ denotes the ensemble average and $\delta(\mathbf{z}) = \prod_{i=1}^n \delta(z_i)$ is the Dirac delta function.
A set of meaningful CVs that properly characterizes the mechanism of the transition between $A$ and $B$ will yield a committor distribution with a single, sharp peak for any value of $\mathbf{q}$. If, for example, $\mathbf{q}=\mathbf{q}^{*}$ coincides with the transition state, the committor distribution should be narrowly peaked around $p_B = 0.5$.  In contrast, a wide spread, multimodal committor distribution is a clear sign of a poor choice of CVs as reaction coordinates.

A first indication of the quality of collective variables is provided by the 
the averaged committor~\cite{bolhuis2011relation} 
\begin{equation}
\label{eq:averaged_committor}
\bar{p}_B (\mathbf{q}) = \frac{\langle p_B(\mathbf{r}) \delta\left( \mathbf{q}(\mathbf{r}) - \mathbf{q} \right) \rangle}{\langle \delta\left( \mathbf{q}(\mathbf{r}) - \mathbf{q} \right) \rangle}
\end{equation}
which is given by integrating over $p_B$ in Eq.~\eqref{eq:committor_dist}. The averaged committor needs to monotonically increase along a set of relevant CVs and can be used to propose tentative transition state ensembles with $\bar{p}_B (\mathbf{q}) = 0.5$ that can then be further scrutinized by analyzing the committor distribution.

\subsection{Transition Path Sampling}
Due to the rare event nature of nucleation processes an advanced simulation technique is required to study the transition between states $A$ and $B$, in this case the liquid and the solid. Here, we employ transition interface sampling,~\cite{van2005elaborating} a variant of transition path sampling,~\cite{bolhuis2002transition,dellago2002transition} that utilises a progress parameter $\lambda$ to define a set of interfaces as hypersurfaces between the two metastable states. 
For each interface, $\lambda_i$, an ensemble of dynamical trajectories is harvested by a Monte Carlo (MC) sampling in path space. The individual path ensembles for each interface can be combined into a complete path ensemble by reweighting the path probabilities. This reweighted path ensemble (RPE)~\cite{rogal2010reweighted} comprises the unbiased dynamics of the system in the full phase space and allows for a direct calculation of various rare event properties~\cite{bolhuis2011relation}. 
In particular, the averaged committor in Eq.~\eqref{eq:averaged_committor} can be projected from the RPE onto any set of collective variables~\cite{rogal2010reweighted}
\begin{equation}
\label{eq:rpe_project}
\bar{p}_B(\mathbf{q}) = \frac{\int\mathcal{D} \mathbf{x}^L \mathcal{P}[\mathbf{x}^L] \mathbbm{1}_B (\mathbf{x}_L) \sum_{k=0}^L \delta(\mathbf{q}(\mathbf{x}_k)-\mathbf{q})}{\int\mathcal{D} \mathbf{x}^L \mathcal{P}[\mathbf{x}^L] \sum_{k=0}^L \delta(\mathbf{q}(\mathbf{x}_k)-\mathbf{q})} \quad ,
\end{equation}
where $\mathcal{P}[\mathbf{x}^L]$ is the reweighted path ensemble, $\int\mathcal{D} \mathbf{x}^L$ denotes the integral over all phase space trajectories of all lengths $L$, $\mathbf{x}^L = \{\mathbf{x}_0, \dots, \mathbf{x}_L\}$, and $\mathbbm{1}_B (\mathbf{x}_L)$ is an indicator function that is one if the last slice of the trajectory, $\mathbf{x}_L$, is in state $B$ and zero otherwise.
To a first approximation, the averaged committor projected from the RPE in Eq.~\eqref{eq:rpe_project} can be used to examine the correlation of various CVs with the committor and identify possible transition state ensembles with $\bar{p}_B(\mathbf{q}^{*}) \approx 0.5$.
By combining the RPE with a maximum likelihood estimation~\cite{peters2006obtaining} it is even possible to achieve a quantitative comparison of the quality of different CVs as reaction coordinates.~\cite{peters2007extensions,lechner2010nonlinear,diaz2018maximum}


\subsection{Collective Variables for $\text{Ni}_3\text{Al}$}
\label{subsec:cv}

To study the nucleation mechanism in $\text{Ni}_3\text{Al}$ we consider a set of 22 different collective variables that comprises parameters concerning the size, crystal structure, and chemical species of the growing solid cluster.

\subsubsection{Size of the Largest Solid Cluster}

A commonly used CV that is often comparable to the radius of the growing nucleus in CNT is the number of particles in the largest solid cluster, $n_s$.
Solid and liquid particles are distinguished based on the Steinhardt bond order parameters.~\cite{steinhardt1983bond}  We use two criteria to identify solid particles, the first is counting the number of solid bonds of each atom $i$ by evaluating the correlation with its neighbors $j$,~\cite{ten1995numerical} $s_{ij}=\sum_{m=-6}^{6}q_{6m}(i)q^{*}_{6m}(j)$, where $q_{6m}$ are the complex vectors calculated from the spherical harmonics with $l=6$.~\cite{steinhardt1983bond}  If $s_{ij} > 0.5$, the connection between $i$ and $j$ is considered as a solid bond.  
The second criterion evaluates the average of the correlation over all neighbors $N_b(i)$ of atom $i$~\cite{bokeloh2011nucleation}, $\langle s_{i} \rangle= 1/N_b(i) \sum_{j=1}^{N_{b}(i)}s_{ij}$, which improves the identification of solid- and liquid-like particles at the solid-liquid interface.
If the number of solid bonds is larger than 7 and $\langle s_{i} \rangle > 0.6$, an atom is considered as solid. The number of atoms in the largest solid cluster, $n_s$, is then determined by a clustering algorithm.

Furthermore, we define the number of skin $n_\text{sk}$ and core $n_c$ atoms. Skin atoms are solid particles in the largest cluster that have at least one liquid neighbor. Core atoms are part of the largest solid cluster and do not have any liquid neighbors. Correspondingly, the sum of skin and core atoms yields the largest solid cluster size, that is $n_{s}=n_\text{sk}+n_{c}$.


\subsubsection{Crystal Structure Identification}

The local crystalline structure around each atom is determined by employing the averaged version of the Steinhardt bond order parameters~\cite{lechner2008accurate} with $l=4,6$, that is $\bar{q}_{4}(i)$ and $\bar{q}_{6}(i)$. A reference map in the $\bar{q}_4-\bar{q}_6$ space~\cite{lechner2011role,lechner2011reaction,diaz2017atomistic} was calculated for $\text{Ni}_3\text{Al}$ including probability distributions for fcc, bcc, hexagonal closed-packed (hcp), and liquid structures (see Supplementary  Material Figure S1). An atom $i$ is assigned to the structure with the highest probability in the reference map for the corresponding $\bar{q}_{4}(i) - \bar{q}_{6}(i)$ value. If all probabilities are less than $10^{-5}$, the particle is labeled \emph{undefined}.

Based on the local structure identification, we define additional collective variables comprising the fraction of fcc, bcc, hcp, and undefined atoms in the largest solid cluster, $\frac{n_s^\text{fcc}}{n_s}$, $\frac{n_s^\text{bcc}}{n_s}$, $\frac{n_s^\text{hcp}}{n_s}$, $\frac{n_s^\text{un}}{n_s}$, and in the core of the largest cluster, $\frac{n_c^\text{fcc}}{n_c}$, $\frac{n_c^\text{bcc}}{n_c}$, $\frac{n_c^\text{hcp}}{n_c}$, $\frac{n_c^\text{un}}{n_c}$, respectively.


\subsubsection{Global Crystallinity within the Solid Cluster}
\label{subsec:crystallinity}

In addition to a local structure identification we consider a global orientational order parameter~\cite{ten1995numerical,jungblut2016pathways}, $Q_{6}^\text{cl}=(\frac{4\pi}{13}\sum_{m=-6}^{6}|\frac{\sum_{i=1}^{N}N_{b}(i)q_{6m}(i)}{\sum_{i=1}^{N}N_{b}(i)}|^{2})^\frac{1}{2}$, 
where the sum runs over all particles in the largest solid cluster~\cite{moroni2005interplay}. 
This CV measures the degree of crystallinity within a solid cluster: solid clusters that are more compact and well ordered show  high values of $Q_{6}^\text{cl}$ and vice versa. In $\text{Ni}_3\text{Al}$, perfect fcc bulk exhibits the highest crystallinity value of all structures with $Q_{6}^\text{cl}=0.48$.

\subsubsection{Chemical Composition and Short-Range Order}

In binary systems, collective variables assessing the chemical composition and the chemical order are also of interest, in particular with respect to phase separation and order-disorder transitions.
To monitor deviations from the ideal 3:1 ratio of nickel and aluminum in $\text{Ni}_3\text{Al}$ we determine the relative amount of Al in the liquid  $\frac{n_\text{liq}^\text{Al}}{n_\text{liq}}$, the largest solid cluster $\frac{n_{s}^\text{Al}}{n_{s}}$, the skin atoms $\frac{n_\text{sk}^\text{Al}}{n_\text{sk}}$, and the core atoms $\frac{n_{c}^\text{Al}}{n_{c}}$, respectively. 
Likewise, the fraction of Al for specific crystalline phases in the largest solid cluster is evaluated with  $\frac{n_{s}^\text{fcc-Al}}{n_{s}^\text{fcc}}$, $\frac{n_{s}^\text{bcc-Al}}{n_{s}^\text{bcc}}$, $\frac{n_{s}^\text{hcp-Al}}{n_{s}^\text{hcp}}$, and $\frac{n_{s}^\text{un-Al}}{n_{s}^\text{un}}$.

The chemical order of Ni and Al is determined by a short-range order parameter~\cite{atanasov2009multi} that was suggested for measuring SRO in nanostructures, such as small clusters. The SRO parameter is a normalized pair-correlation function with spin-like variable $S_i$ for the site occupancy, that is $S_i = 1$ and $-1$ for atomic species A and B, respectively.
For a binary alloy, the SRO parameter of site $i$ for the $n$-th neighbor shell is defined as 
\begin{equation}
\label{eq:sro}
\text{SRO}_n= \frac{\langle {S_{i}}{S_{i+n}}\rangle - \langle S_i \rangle^{2}}{1-\langle S_i \rangle^2}
\end{equation}
where $\langle S_i \rangle = 2x-1$ and $x$ is the overall fraction of the chemical element occupying site $i$, i.e. $x_\text{Ni} = 0.75$ and $x_\text{Al} = 0.25$ for $\text{Ni}_3\text{Al}$.
If the lattice sites are randomly occupied by Ni and Al, the correlation parameter is SRO$_n=0$. For the chemically ordered $\text{L1}_{2}$ phase, SRO$_1=-1/3$ for the first  and SRO$_2=1$ for the second neighbor shell, respectively. Since $\text{L1}_{2}$ is the most stable phase in $\text{Ni}_3\text{Al}$,~\cite{xu1990phase,massalski1992binary} we use these SRO parameter values as reference to quantify the chemical order of crystalline embryos that emerge within the melt.   


\subsection{Computational Details}
\label{subsec:compdetail}

All simulations were performed in a cubic box with 6912 atoms, corresponding to a $(12\times 12\times 12)$ fcc supercell, and a fixed composition of 75~at.\% Ni and 25~at.\% Al. The interatomic interactions were modeled with an embedded atom method (EAM) potential for Ni-Al~\cite{purja2009development} that was shown to capture the stability of various stable and metastable phases across the phase diagram. 
The EAM potential gives a melting temperature for $\text{L1}_2$ ordered $\text{Ni}_3\text{Al}$ of approximately $T_{m}=1678$~K~\cite{purja2009development}, which agrees well with the experimental values of $T_{m}=1645$~K~\cite{massalski1992binary}. 
Dynamical trajectories were created by molecular dynamics (MD) using the simulation package {\scshape lammps}~\cite{plimpton1995fast} in the isothermal-isobaric (NPT) ensemble with a Nos{\'e}-Hoover thermostat and barostat. The damping time regulating the temperature and pressure were set to 0.05~ps and 0.5~ps. Only the volume of the simulation box was allowed to change while the shape was kept cubic. The pressure and temperature were $P=0$~bar and $T=1342$~K, which corresponds to an undercooling of 20\% with respect to the melting temperature of $\text{Ni}_3\text{Al}$ given by the Ni-Al EAM.~\cite{purja2009development}. The integration timestep was set to $\Delta t= 1$~fs, and three-dimensional periodic boundary conditions were applied in all MD simulations.

For the TIS simulations, we adopted a python wrapper~\cite{tpswrapper} combined with {\scshape lammps} to perform the MC sampling of the MD trajectories. As progress parameter $\lambda$ the size of the largest solid cluster $n_s$ was used. In total, there are 21 interfaces  with $\lambda = \{15,25,38,50,63,80,100, 125,140,170,\\ 200,230,260,300,320,330,350,400,450,475,500\}$. The positions of the interfaces are chosen such that there is at least 10\% overlap in the crossing histograms between neighboring interfaces. The first interface $\lambda_0 = 15$ marks the boundary of the liquid state and the last one $\lambda_{20} = 500$ of the solid state. 
For cluster sizes $n_s \geq 500$ the system is fully committed to the solid state and complete solidification occurs. The path ensemble was harvested with replica exchange TIS (RETIS)~\cite{van2007reaction,bolhuis2008rare} with 45\% shooting moves, 45\% exchange moves, 10\% exchange between the forward and backward ensembles. For each interface 1800 MC moves were performed where trajectories were collected after 5 decorrelation steps, generating 360 paths in each interface ensemble. All atomic positions and velocities were recorded at 0.2~ps time intervals.

To compute the committor distribution $P(p_B|_ \mathbf{q})$ in Eq.~\eqref{eq:committor_dist} a minimum number of 50 configurations with a given value of $\mathbf{q}$ was randomly chosen  from the path ensemble. For each configuration 100 MD simulations were initiated with Maxwell-Boltzmann distributed velocity at $T=1342$~K and it was monitored if the trajectory entered the solid ($n_s \geq 500$) or liquid ($n_s \leq15$) state first, yielding the committor $p_B(\mathbf{q})$. All $p_B(\mathbf{q})$ values are converged to within $\pm0.05$.

\section{Results}
\label{sec:results}

\subsection{Multiple Reaction Channels During Nucleation}
\label{subsec:fccbcccompete}
\begin{figure}
\centering
\begin{subfigure}{0.45\linewidth}
\caption{\label{fig:sub1dprojection}}
  \includegraphics[width=0.9\linewidth]{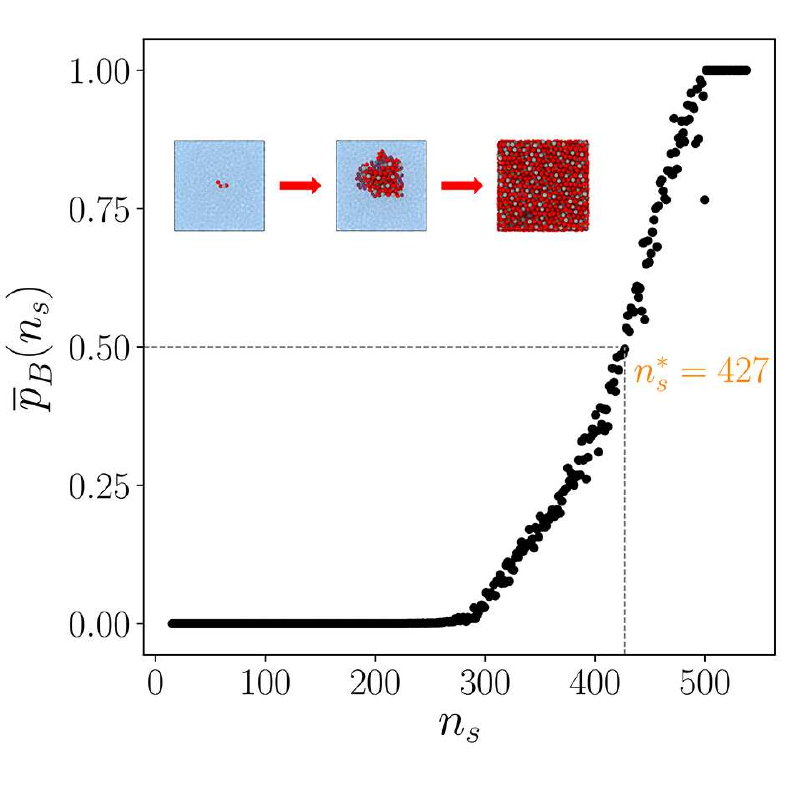}
\end{subfigure}
\hfill
\begin{subfigure}{0.45\linewidth}
\caption{\label{fig:sub1dcommittordistribution}}
  \includegraphics[width=0.9\linewidth]{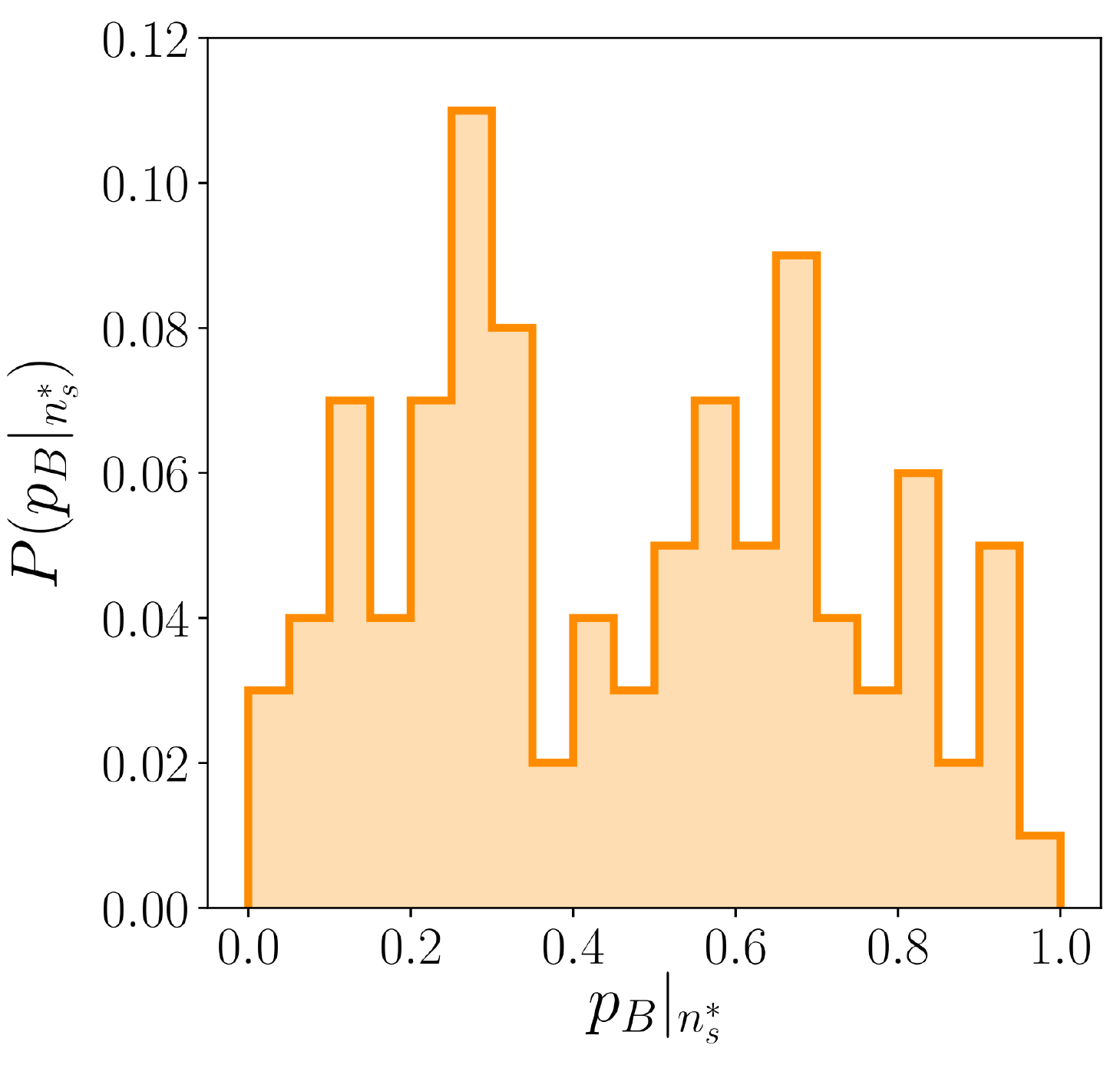}
\end{subfigure}
\caption{(a) Averaged committor $\bar{p}_{B}(n_{s})$ projected from the RPE. $\bar{p}_{B}(n_{s})$ and $n_{s}$ are closely correlated, the assumed transition state is at $n_s^* = 427$. The inset illustrates nucleation and crystallization in $\text{Ni}_3\text{Al}$. The liquid particles, solid atoms of Ni and Al are presented in transparent blue, red, and gray, respectively. (b) Committor distribution $P(p_{B}|_{n_{s}^{*}})$ extracted from 100 transition state configurations with $n_s^* = 427$.\label{fig:pbofns}}
\end{figure}

An important step in analyzing the mechanism of nucleation in $\text{Ni}_3\text{Al}$ is the identification of a suitable reaction coordinate. The size of the largest solid cluster, $n_s$, is a first, intuitive choice, as it was shown to provide a complete description of the nucleation process in many systems, including LJ~\cite{van2008two}, unary metals~\cite{diaz2017atomistic}, and soft core colloids~\cite{lechner2011role}.  
In Figure~\ref{fig:sub1dprojection} the averaged committor projected onto $n_s$ from the RPE is presented. The committor increases monotonically with the typical shape indicating a good correlation between $n_s$ and $\bar{p}_B$. The averaged committor is, however, not sufficient to evaluate the quality of a CV as reaction coordinate. Previous studies of crystal nucleation in pure Ni showed that even though the averaged committor exhibits a good correlation with several CVs such as the size of the crystalline core $n_c$, this parameter is unable to capture the entire nucleation process and the corresponding free energy barrier is significantly underestimated.~\cite{diaz2018maximum}

To provide a quantitative analysis of the quality of $n_{s}$ as RC, we perform a committor analysis with configurations belonging to the apparent TS $n_{s}^{*}=427$, with $\bar{p}_B(n_s^*) = 0.5$.
We collect 100 TS configurations to compute the committor distribution $P(p_B|_{n_s^*})$ according to Eq.~\eqref{eq:committor_dist}.
As shown in Figure~\ref{fig:sub1dcommittordistribution}, the committor distribution does not yield a single narrow peak at $p_{B} \approx 0.5$ that is characteristic of the TS ensemble, but is spread out over the entire range of committor values. The size of the largest cluster $n_s$, therefore, does not yield a good approximation to the RC, despite its reasonable correlation with the averaged committor.

\begin{figure}
\centering
\begin{subfigure}{0.45\linewidth}
\caption{\label{fig:subns427fractions}}
  \includegraphics[width=0.9\linewidth]{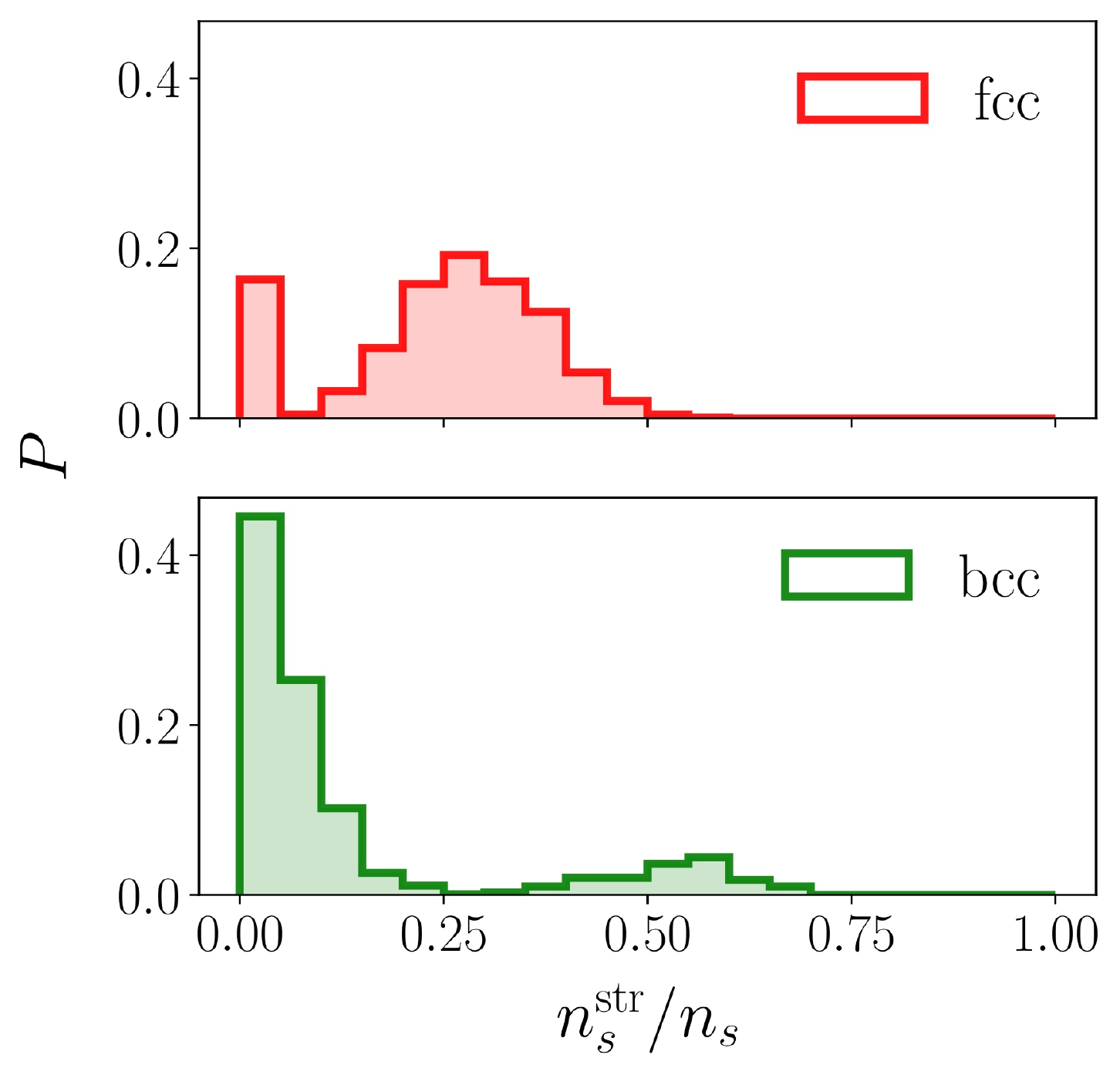}
\end{subfigure}
\hfill
\begin{subfigure}{0.45\linewidth}
\caption{\label{fig:subns427relation}}
  \includegraphics[width=0.9\linewidth]{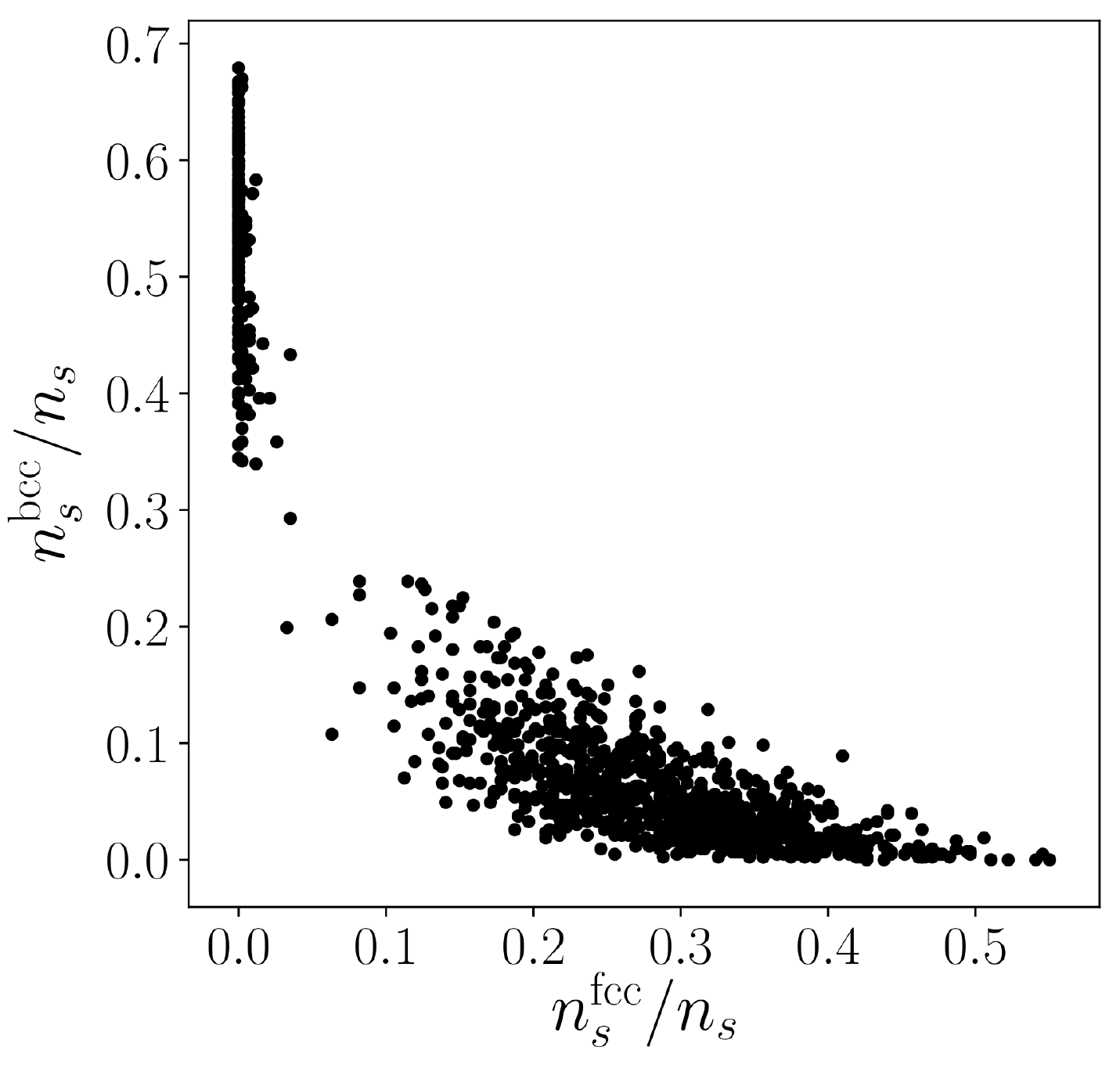}
\end{subfigure}
\caption{(a) Distribution of the phase fractions of fcc (red) and bcc (green) in clusters with a presumed critical size $n_{s}^{*}=427$ computed from 1087 configurations in the path ensemble. (b) Correlation between the phase fractions of fcc and bcc.\label{fig:ns427str}}
\end{figure}

To understand why this widely and successfully used CV fails to describe the nucleation mechanism in $\text{Ni}_3\text{Al}$, we analyze the structural composition of the growing clusters. Unlike in pure Ni or Al where the core of the solid clusters is dominated by fcc~\cite{diaz2017atomistic,desgranges2007molecular}, in $\text{Ni}_3\text{Al}$ we find a mixture of fcc, bcc, and random stacking of hcp (rhcp). More specifically, two nucleation pathways seem to exist that lead to the formation of fcc and bcc phases, respectively.
This can, for example, be seen from the structural composition of configurations at the supposed TS with $n_{s}^{*}=427$. The distribution of the phase fractions of fcc and bcc in the presumed critical clusters in Figure~\ref{fig:subns427fractions} is approximately bimodal: $\frac{n_s^\text{fcc}}{n_s}$ has two peaks around 0.3 and 0.0, and  $\frac{n_s^\text{bcc}}{n_s}$ at 0.6 and 0.0, respectively. Moreover, the formation of fcc and bcc in the growing clusters appears to be mutually hindered. As shown in Figure~\ref{fig:subns427relation} the amount of fcc (bcc) substantially increases only in clusters with a rather low bcc (fcc) content. Smaller (pre-critical) and larger (post-critical) clusters exhibit the same accumulation of either fcc or bcc, which indicates that there is a competition between fcc and bcc during nucleation and growth leading to two separate reaction channels. Similar to the nucleation in unary metals~\cite{diaz2017atomistic}, we observe the formation of pre-structured liquid~\cite{diaz2018maximum} and rhcp for both nucleation pathways.

The structural composition of the small nuclei directly effects the formation of specific polymorphs in the final bulk phase. MD simulations started from presumed TS configurations with clusters that were predominantly composed of fcc or bcc showed that the solidified bulk phase inherits the structure of the initial nuclei. The selection of different polymorphs thus takes place in the early stages of nucleation triggered by the structural composition of the small, initial clusters.

The existence of competing nucleation pathways was also found for the crystallization in LJ systems~\cite{moroni2005interplay,beckham2011optimizing}, methane hydrates~\cite{berendsen2019unbiased} and mixtures of oppositely charged colloidal particles~\cite{sanz2007evidence,peters2009competing}. During nucleation in a LJ liquid the critical clusters can be either small, compact and mostly fcc, or large, loosely-packed and more bcc-like~\cite{moroni2005interplay}. In methane hydrates two nucleation pathways were identified at moderate undercooling, one towards the thermodynamically stable crystalline hydrate, and the other resulting in a metastable amorphous phase.~\cite{berendsen2019unbiased}
In the mixture of oppositely charged colloidal particles, competing nucleation pathways were found to coexists in a broad reaction channel with selection occurring near the barrier top leading to different bulk phases of charge-disordered fcc and CsCl-ordered bcc~\cite{peters2009competing}.
In $\text{Ni}_3\text{Al}$ the thermodynamically stable phase is $\text{L1}_2$ ordered fcc, but there are several chemically ordered and disordered metastable phases including bcc and hcp.~\cite{xu1990phase}  
Experimental studies on crystallization in Ni-Al alloys indeed revealed a variety of possible transitions between ordered and disordered fcc and bcc phases, depending on the composition as well as the applied undercooling.~\cite{assadi1998kinetics} This supports our findings of a competition between the nucleation and growth of either fcc or bcc in $\text{Ni}_3\text{Al}$. The relative probability of these two nucleation mechanisms could, however, not be assessed within the current study.  Depending on the composition and undercooling the path probability density in the two reaction channels varies resulting in a weighted contribution of the different paths to the overall nucleation process.

Since the nucleation mechanism strongly depends on the competition between different crystal structures in the growing cluster, it becomes clear why a CV that only measures the size, $n_s$, is not sufficient as reaction coordinate.  To obtain a more suitable description we add in a next step a measure of the crystallinity to the reaction coordinate.

\subsection{Importance of Crystallinity}
\label{subsec:2dRCcommitanalysis}

\begin{figure}
\centering
\begin{subfigure}{0.525\linewidth}
\centering
\caption{}
\label{fig:sub2dprojection}
  \includegraphics[height=6cm,width=0.9\linewidth]{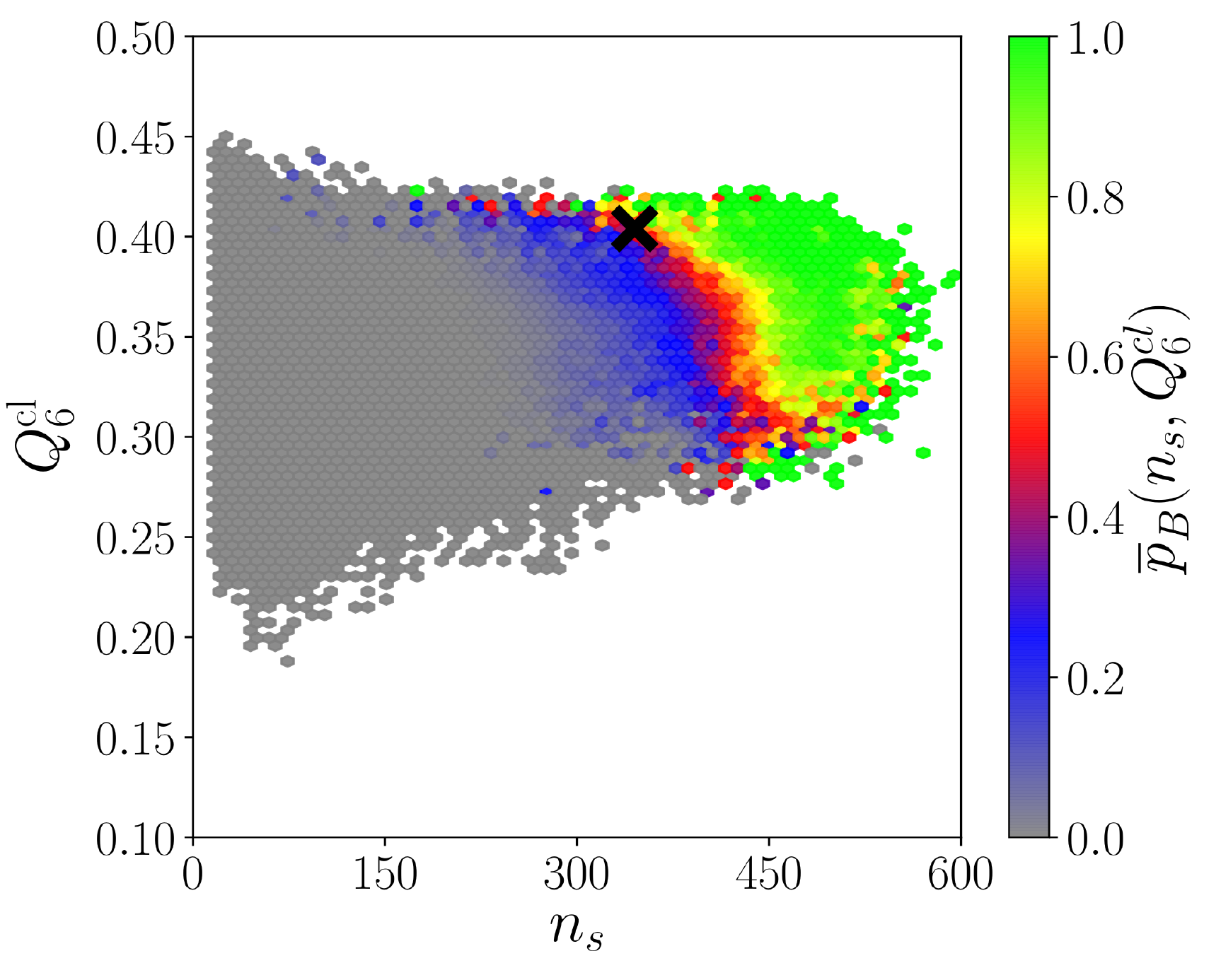}
\end{subfigure}
\hfill
\begin{subfigure}{0.445\linewidth}
\centering
\caption{}
\label{fig:subpBhistons320Q6390}
  \includegraphics[height=6cm,width=0.9\linewidth]{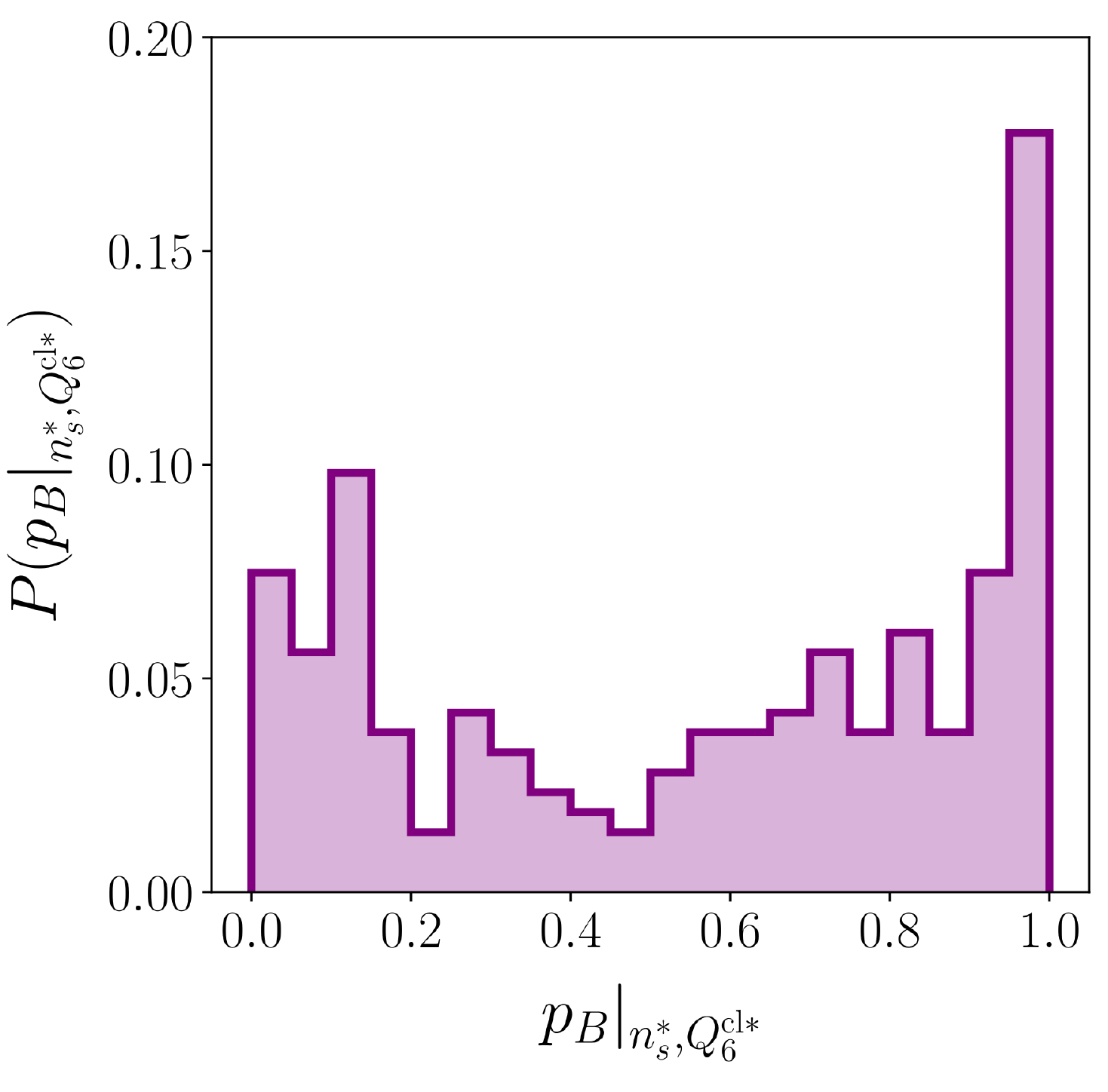}
\end{subfigure}
\caption{
(a) Averaged committor $\bar{p}_B (n_s, Q_6^\text{cl})$ projected from the RPE. The TS region $\bar{p}_B(n_s,Q_6^\text{cl}) \approx 0.5$ is colored in red. (b) Committor distribution $P(p_B|_{ n_{s}^*,Q_{6}^\text{cl*}})$ obtained from 214 (assumed) TS configurations with $n_s^* = 320\pm5$ and $Q_6^\text{cl*} \geq 0.39$. The TS configurations were chosen from the region of $\bar{p}_B (n_s, Q_6^\text{cl})$ marked by a black cross. }
\label{fig:2dpbs}
\end{figure}

The global crystallinity $Q_6^\text{cl}$ (section~\ref{subsec:crystallinity}) was suggested as an important parameter in the description of nucleation in a LJ liquid where likewise the nucleation mechanism proceeded via several different pathways.~\cite{moroni2005interplay}
Here, we project the averaged committor from the RPE on both, the size of the largest cluster $n_s$ and the crystallinity $Q_6^\text{cl}$, shown in Figure~\ref{fig:sub2dprojection}.
The TS region with $\bar{p}_{B}(n_{s},Q_{6}^\text{cl}) \approx 0.5$ exhibits a non-linear dependence on the two CVs indicating that at least a two-dimensional RC is required to capture the nucleation mechanism.
The presumed TS (red area in Figure~\ref{fig:sub2dprojection}) covers a fairly wide range of critical cluster sizes, $300 < n_s^* < 460$, and crystallinity values, $0.28 < Q_6^\text{cl*} < 0.42$. Critical clusters can be either small with high crystallinity ($n_{s} \approx 320$, $Q_{6}^\text{cl} \geq 0.39$), composed of predominantly fcc, negligible bcc, and some rhcp, or they can be large with lower crystallinity ($n_{s} \approx 458$, $Q_{6}^\text{cl} \leq 0.33$), consisting of mainly rhcp with comparable amounts of fcc and bcc (see Supplementary Material Figure S2).
The variety of cluster sizes and crystallinity in the TS ensemble further corroborates the existence of multiple reaction channels as well as the necessity of a multi-dimensional RC.

To evaluate the quality of our two-dimensional RC model we performed a committor analysis on the presumed TS ensemble for both small/high-crystallinity and large/low-crystallinity clusters. The committor distribution $P(p_B|_{n_s^*, Q_6^\text{cl*}})$ computed from 214 configurations with small, compact nuclei satifying $n_s^* = 320\pm5$ and $Q_6^\text{cl*} \geq 0.39$ is shown in Figure~\ref{fig:subpBhistons320Q6390}. The committor values still extend over the entire range, but there is a slight increase in the probability distribution for low ($p_B \approx 0.0$) and high ($p_B \approx 1.0$) values. For large/low-crystallinity clusters of $n_{s}^*=458$, $Q_{6}^\text{cl*} \leq 0.33$ (54 configurations), the committor distribution $P(p_B|_{n_s^*, Q_6^\text{cl*}})$ is as well spread over the whole range of committor values with a slight increase in the range of $0.2-0.6$ (Supplementary Material S3a).  
Even though the two-dimensional RC model somewhat improves the description of the nucleation mechanism, and both $n_s$ and $Q_6^\text{cl}$ are important characteristics, there is still an additional component missing. 


\subsection{Chemical Short-Range Order in the Growing Cluster}
\label{subsec:SRO}
As a measure of how much an additional CV might improve the RC we analyze their respective correlation with the committor for fixed values of $n_s$ and $Q_6^\text{cl}$.  Specifically, we evaluate 20 different parameters that characterize the crystal structure as well as the chemical composition and order as introduced in section~\ref{subsec:cv}. For small, compact clusters ($n_s^* = 320\pm5, Q_6^\text{cl*} \geq 0.39$) all CVs that provide additional information about the structure, that is the number of core ($n_c$) and skin ($n_\text{sk}$) atoms, and the fraction of fcc, bcc, hcp, and undefined in the largest cluster and in the core, do not show any significant correlation with the committor (see Supplementary Material Figure S4).  This indicates that the global crystallinity already captures the important structural characteristic needed to distinguish different pathways during nucleation and growth in $\text{Ni}_3\text{Al}$.
Furthermore, the chemical composition does not play any role as additional parameter in the RC. The amount of Ni and Al is approximately constant in the liquid as well as in the solid clusters, also when further separated into core and skin atoms or the different crystalline phases (see Supplementary Material Figure S4). Similar uncorrelated behavior of structural CVs and chemical composition with the committor is also observed for large/low-crystallinity clusters ($n_{s}^*=458$, $Q_{6}^\text{cl*} \leq 0.33$).

The chemical short-range order, however, does exhibit a strong correlation with the committor for both the first and second nearest neighbour shell. 
\begin{figure}
\centering
\includegraphics[width=8.5cm]{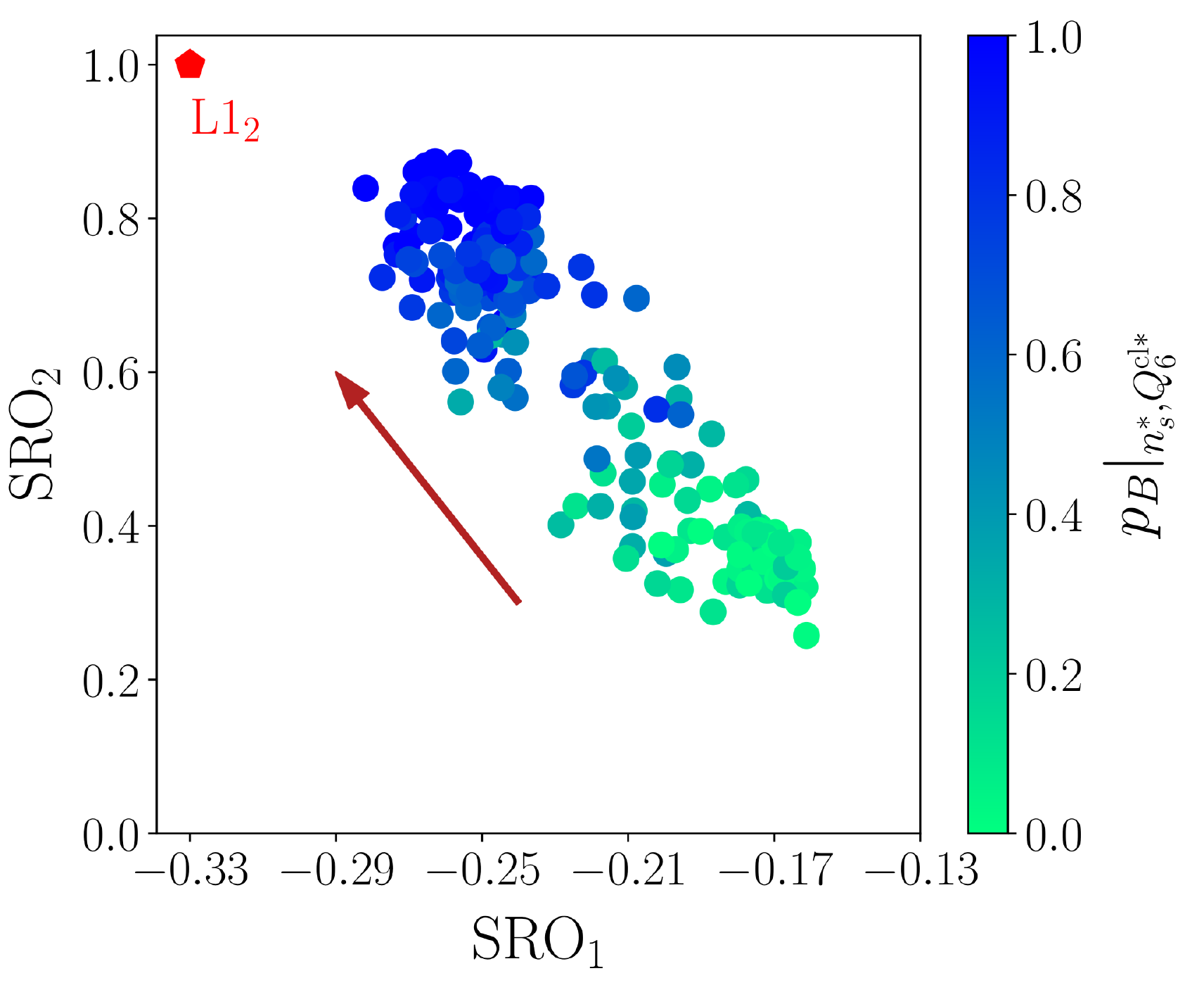}
\caption{\label{fig:OPscorrelatepB} Committor values of assumed TS configurations for small,compact clusters with $n_{s}^* = 320\pm5$ and $Q_{6}^\text{cl*} \geq 0.39$ as a function of the chemical short-range order. The arrow indicates an increase in SRO values towards $\text{L1}_2$ marked by a red pentagon.}
\end{figure}
In Figure~\ref{fig:OPscorrelatepB} the committor values of all configurations with small, compact clusters are plotted as a function of SRO$_1$ and SRO$_2$. Clearly, the committor increases with increasing SRO towards an $\text{L1}_2$ ordering with SRO$_1 = -1/3$ and SRO$_2 = 1.0$.  In these small clusters, the SRO of the first and second nearest neighbours are not independent, confirming the clear correlation between an $\text{L1}_2$ ordering in the clusters and an increase in $p_B$.  This suggests that the chemical SRO does indeed provide additional information needed for an appropriate description of the nucleation process. For large clusters with low-crystallinity, correlations of SRO$_1$ and SRO$_2$ with the commitor are significantly reduced due to the clusters polymorphic composition, where competing rhcp, bcc, and fcc phases make the definition of SRO ambiguous (see Supplementary Material Figure S3b). In the following we only focus on the small/high-crystallinity clusters to evaluate the effect of chemical ordering on the RC since the dominant phase fraction of fcc in the core avoids the noise in SRO values that emerges from competing phases.

\begin{figure}
\includegraphics[width=8.5cm]{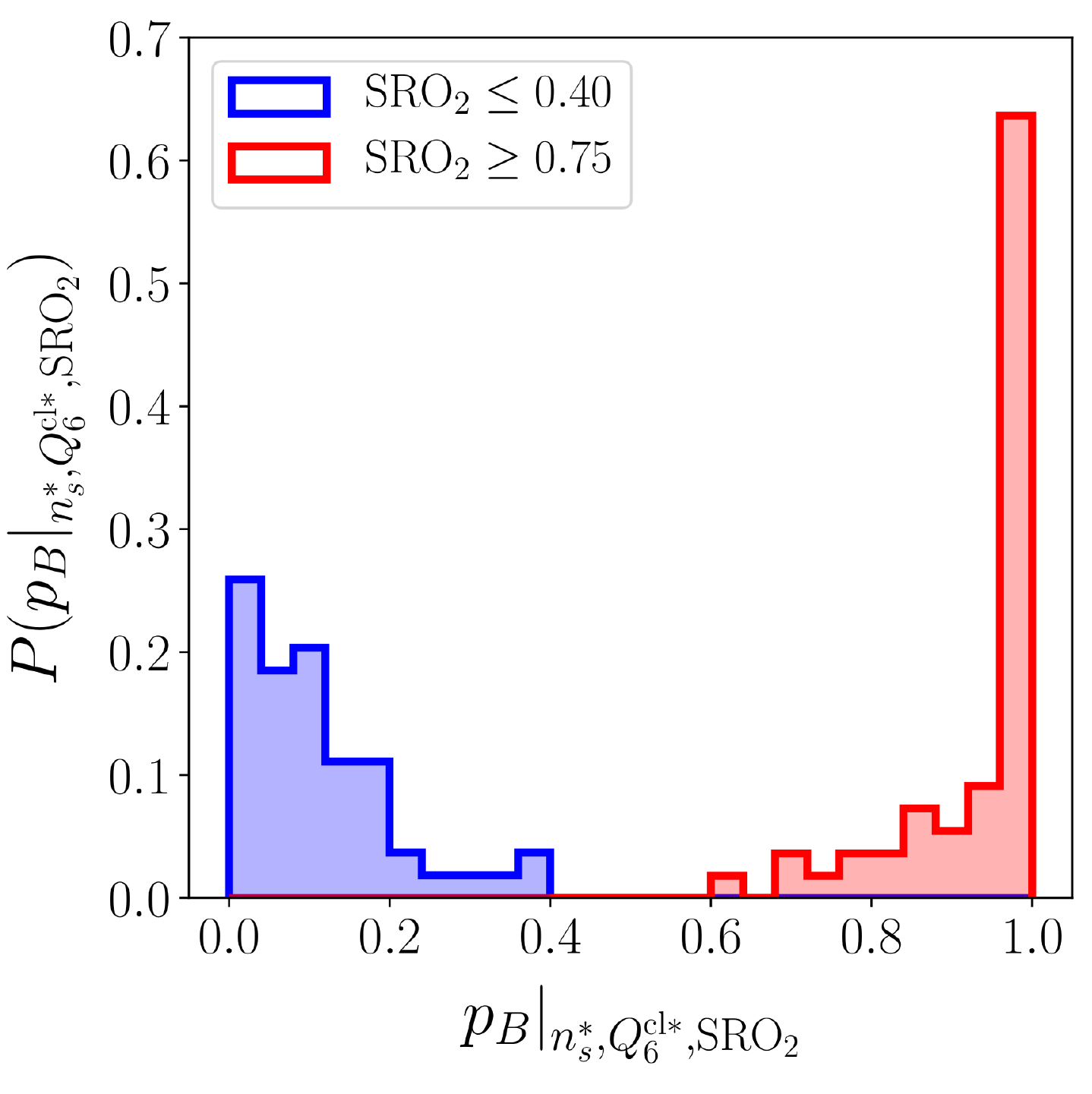}
\caption{Committor distributions $P(p_B|_{n_s^*, Q_6^\text{cl*}, \text{SRO}_2})$ for two sets of configurations with the same size $n_s^* = 320\pm5$ and crystallinity $Q_6^\text{cl*} \geq 0.39$, and low SRO$_2 \leq 0.4$ (blue) and high SRO$_2 \geq 0.75$ (red), respectively. The distribution for  clusters with a chemical ordering close to $\text{L1}_2$ is peaked around $p_B \approx 1.0$ and for nearly disordered clusters around $p_B \approx 0.0$.\label{fig:highlowchemicalsropb}}
\end{figure}

To confirm our hypothesis that SRO significantly improves the description of the nucleation mechanism, we perform a committor analysis within the three-dimensional RC space including the size, crystallinity, and short-range order of the largest solid cluster.  Specifically, SRO$_2$ is added as third component to the RC. 
We define two sets of configurations with small, compact clusters ($n_{s}^* = 320\pm5$, $Q_{6}^\text{cl*} \geq 0.39$): one with high chemical order, SRO$_2 \geq 0.75$ (55 configurations), and the other with low chemical order, SRO$_2 \leq 0.40$ (54 configurations).
The corresponding committor distributions are shown in Figure~\ref{fig:highlowchemicalsropb}. The two sets result in clearly separated committor distributions that are sharply peaked at $p_B \approx 1.0$ for nuclei with an ordering tendency towards $\text{L1}_2$, and $p_B \approx 0.0$ for clusters showing more chemical disorder, respectively.
The chemical SRO therefore stabilizes the small, compact clusters, so that they are already beyond the critical nucleus size and continue to grow into the solid bulk phase, whereas the more chemically disordered clusters of the same size and crystallinity dissolve again into the liquid phase.
The single peak in both committor distributions suggests that the three-dimensional representation of the RC considering size, crystallinity, and chemical order can rigorously capture the nucleation mechanism.
%


\subsection{Nucleation Mechanism and Growth in $\text{Ni}_3\text{Al}$}

While in many simple liquids and unary metals the size of the largest growing cluster is a good approximation to the RC, the nucleation mechanism in $\text{Ni}_3\text{Al}$ with several distinct pathways and the formation of various chemically ordered and disordered crystal phases requires a multi-dimensional RC that can account for and differentiate between these aspects, as evident from the previous section.
The importance of the different features and, correspondingly, the probability of different nucleation pathways will also depend on the environmental conditions, such as pressure and undercooling.

In the current study at 20\% undercooling the chemical SRO plays a key role in stabilizing growing clusters with high crystallinity. This originates in the strong ordering tendency of the bulk phases where for $\text{Ni}_3\text{Al}$ the cohesive energy of $\text{L1}_2$ ($E^\text{coh}_{\text{Ni}_3\text{Al} - \text{L1}_2} = -4.63$~eV/atom) is about 100~meV/atom lower than the one of disordered fcc ($E^\text{coh}_{\text{Ni}_3\text{Al} - \text{fcc}} = -4.51$~eV/atom), calculated using the Ni-Al EAM potential.~\cite{purja2009development} 
At finite temperatures the free energy difference becomes smaller which is mainly due to configurational entropy whereas vibrational entropy differences appear to be small between the two phases.~\cite{vandewalle98}
Still, experimental as well as theoretical studies indicate that the ordered $\text{L1}_2$ phase remains thermodynamically stable up to the melting temperature.~\cite{cahn1987order,Michelon10}
Comparing the stability of the small, compact cluster ($n_{s}^* = 320$, $Q_{6}^\text{cl*} \geq 0.39$) with high (SRO$_2 \geq 0.75$) and low (SRO$_2 \leq 0.4$) short-range order we find that the average potential energy of the ordered clusters is approximately 40~meV/atom lower than for disordered ones. At smaller undercoolings we expect the effect to be stronger than at very large undercoolings due to the competition between thermodynamic and kinetic factors. Close to the melting temperature the nucleation barrier is large and the corresponding nucleation rate is low. In addition, the diffusion of atoms in the melt is fast. During the formation of solid nuclei the atoms thus have enough time and are mobile enough to rearrange into the thermodynamically favored phase.
At large undercoolings, however, kinetic effects most likely dominate over thermodynamics.  The nucleation barrier becomes small and the formation of solid nuclei occurs rapidly.  Together with the slower diffusion of atoms this leads to the formation of anti-site defects, and the solid phase inherits the chemical order of the liquid instead of relaxing to the thermodynamically stable phase.  Consequently, the chemical SRO is not expected to significantly impact the nucleation mechanism at large undercoolings.

Even though we observe a strong effect of the chemical SRO on the stability of the growing clusters in the initial stage of the nucleation, the final bulk solids are not $\text{L1}_2$ ordered in our simulations. To understand this we have a closer look at the growth stage during solidification.
During growth the formation of a chemically ordered phase is similarly controlled by a competition between diffusion in the melt and the velocity of the solid-liquid interface.
To evaluate the ordering and interface velocity during growth, simulations were setup within a supercell containing an $\text{L1}_2$ ordered bulk phase in contact with liquid $\text{Ni}_3\text{Al}$ in the [001] direction (see Supplementary Material for further details). After equilibration the system was kept at 1\% undercooling where the liquid phase quickly solidified by growth from the solid-liquid interface. Even at this very small undercooling the estimated interface velocity is high ($v \approx 3$~m/s). As a result the solidified bulk phase does not grow with $\text{L1}_2$ ordering, but exhibits the same chemical SRO as the liquid phase. This disorder trapping~\cite{boettinger1989theory,aziz1994transition} of chemical species is expected whenever the growth velocity exceeds the mobility of atoms in the melt.
Experimental studies~\cite{barth1994rapid} of rapid solidification in Ni-Al alloys as well observed disorder trapping at crystal growth velocity of $v\approx 4$~m/s in $\text{Ni}_3\text{Al}$. The growth of the ordered $\text{L1}_{2}$ phase is predicted to be possible only at growth velocity well below 1~m/s~\cite{assadi1995application,assadi1998kinetics}. 
The chemical SRO thus strongly influences the stability of the growing clusters in the initial stage of nucleation, while it only marginally affects the growth stage due to the large solid-liquid interface velocities accompanied by disorder trapping.

\section{Conclusion}
\label{sec:conclusion}

In summary, we have identified a multi-dimensional RC as a suitable descriptor for the nucleation process in $\text{Ni}_3\text{Al}$ by applying a committor analysis of several CVs on configurations obtained from TIS ensembles. In contrast to unary metals ~\cite{diaz2017atomistic,desgranges2007molecular} and other bimetallic alloys~\cite{desgranges2014unraveling,desgranges2016effect,bechelli2017free,watson2011crystal}, the nucleation mechanism in $\text{Ni}_3\text{Al}$ exhibits particular complexity that arises from the competition of crystalline structures and chemical ordering. Although the size of the largest solid cluster was found to be strongly correlated to the averaged committor and therefore is a key descriptor of the mechanism, our analysis shows that this order parameter is not sufficient as RC model of homogeneous nucleation for this bimetallic compound. Indeed, the structural analysis of the nuclei at fixed critical size obtained from the path ensemble revealed the appearance of various crystalline structures strongly indicating the existence of several nucleation channels.  
Consequently, the RC of the nucleation process in $\text{Ni}_3\text{Al}$ is enhanced by a crystallinity parameter $Q^\text{cl}_6$, similar to what was found by Moroni et al.~\cite{moroni2005interplay} for nucleation in a LJ system. Further analysis of the correlation of CV candidates with the committor unravelled the crucial role of the chemical short-range order to stabilize solid clusters and promote critical fluctuations. 
For the same size and crystallinity, nuclei with increased chemical order towards the $\text{L1}_2$ phase tend to grow and solidify, whereas nearly chemically disordered clusters eventually dissolve in the melt. 
Unlike other bimetallics that form solid solutions,~\cite{desgranges2014unraveling,desgranges2016effect,bechelli2017free,watson2011crystal} the chemical composition of the clusters does not play a role in the nucleation mechanism of $\text{Ni}_3\text{Al}$, as shown by the poor correlation of this CV with the committor and thus its negligible improvement of the RC model. 
The chemical composition remains essentially constant for the growing nuclei and does not impact the nucleation mechanism, whereas the strong ordering tendency of the chemical elements in $\text{Ni}_3\text{Al}$ measured by the SRO parameter strongly enhances the RC.

Our results reveal that a comprehensive description of the nucleation mechanism in $\text{Ni}_3\text{Al}$ requires to take into account the interplay between cluster size, crystallinity, and short-range order. In general, we expect that multi-dimensional models of the RC as obtained from our extensive statistical analysis of the committor and path ensemble in $\text{Ni}_3\text{Al}$ are of key importance in the characterization of nucleation mechanisms in complex alloys.
\section*{supplementary material}
See the Supplementary Material for additional details on $\text{Ni}_3\text{Al}$ $\bar{q_4}-\bar{q_6}$ map for structure identification, structural compositions of presumed critical clusters, committor distribution of large/low-crystallinity critical clusters and correlation between chemical short-range order and committor, correlation of various collective variables with the committor for small/high-crystallinity clusters, rapid growth simulations in $\text{Ni}_3\text{Al}$.
\section*{Data Availability Statement}
The data that support the findings of this study are available from the corresponding author upon reasonable request.
\begin{acknowledgments}
Y. Liang acknowledges the Ph.D fellowship from the International Max Planck Research School for Interface Controlled Materials for Energy Conversion (IMPRS-SurMat). We acknowledge financial support by the Deutsche Forschungsgemeinschaft (DFG) through project 262052203 and project 211503459 (C2 of the collaborative research center SFB/TR 103). 
The authors acknowledge computing time by the Center for Interface-Dominated High Performance Materials (ZGH, Ruhr-Universit{\"a}t Bochum).
\end{acknowledgments}

\bibliography{NiRC}

\end{document}


\begin {CJK*}{UTF8}{gbsn} 
\title[]{Identification of a Multi-Dimensional Reaction Coordinate for Crystal Nucleation in $\text{Ni}_3\text{Al}$: Supplementary Material} 
\author{Yanyan Liang(梁燕燕)}
\email{yanyan.liang@rub.de}
\author{Grisell D\'{i}az Leines}%
\author{Ralf Drautz}
\author{Jutta Rogal}
\email{jutta.rogal@rub.de}
\affiliation{Interdisciplinary Centre for Advanced Materials Simulation, Ruhr-Universit{\"a}t Bochum, 44780 Bochum, Germany}
\date{\today}
\maketitle
\end{CJK*}
\section{$\text{Ni}_3\text{Al}$ $\bar{q}_4 - \bar{q}_6$ Map}

\begin{figure}
\centering
\includegraphics[width=8.5cm]{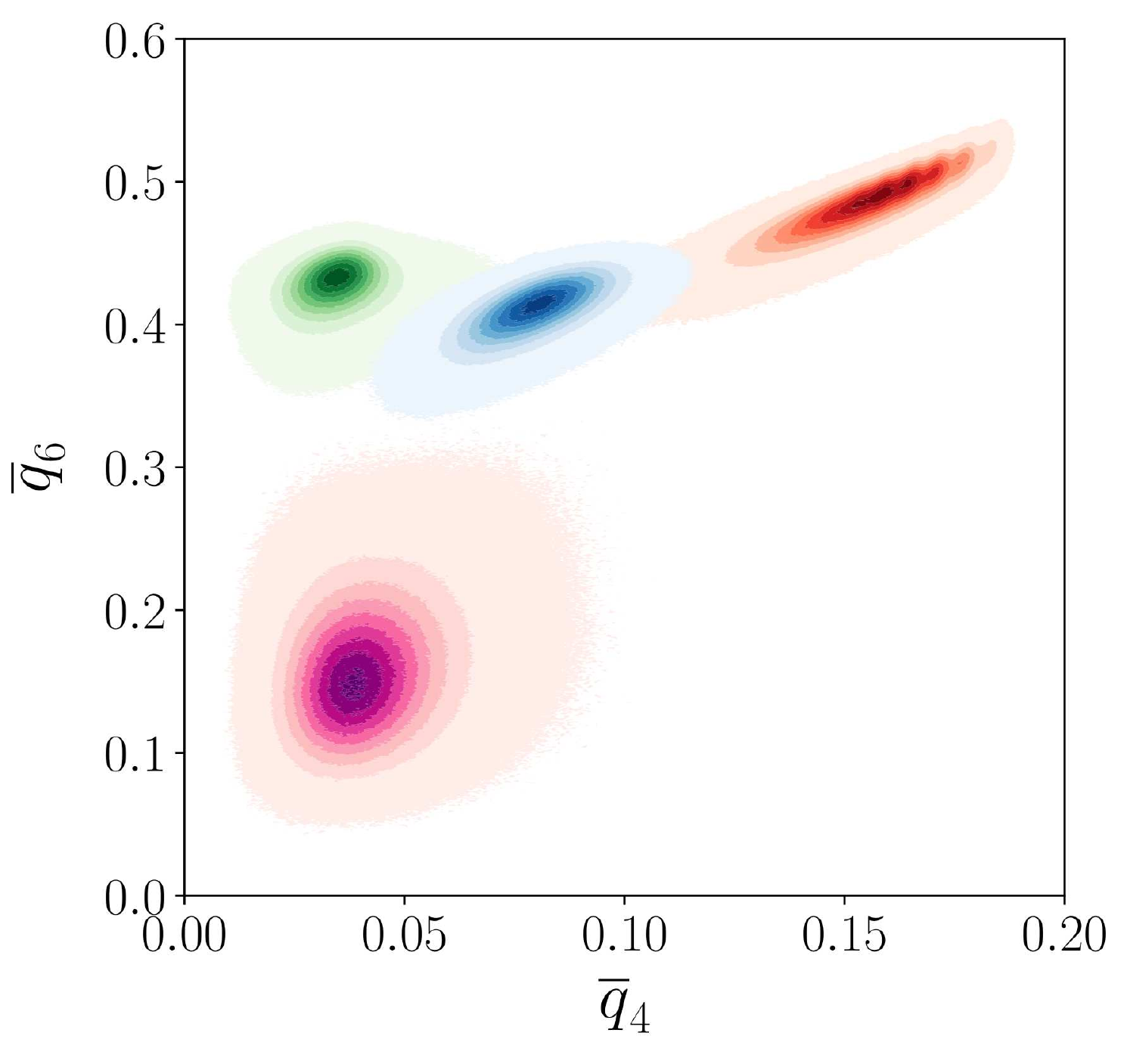}
\caption{ $\bar q_{4}(i)$-$\bar q_{6}(i)$ map for various bulk structures in $\text{Ni}_3\text{Al}$ at 20\% undercooling ($T=1342$~K) including fcc (red), bcc (green), hcp (blue), and liquid (magenta). The color gradient corresponds to the increase of the probability density. }
\label{fig:ni3alq4q6}
\end{figure}

To obtain the $\text{Ni}_3\text{Al}$ $\bar{q}_4-\bar{q}_6$ structural reference map at 20\% undercooling, 5~ns molecular dynamics (MD) trajectories were generated at a fixed composition of 75~at.\% Ni and 25~at.\% Al for different bulk phases. The system sizes are $N=4000$ atoms for fcc, liquid, and $N=3456$ atoms for bcc and hcp. All bulk phases of $\text{Ni}_3\text{Al}$ were created with chemically ordered crystalline phases: fcc - $\text{L1}_2$, bcc - $\text{DO}_3$, and hcp - $\text{DO}_{19}$, respectively. All MD simulations were performed with {\scshape lammps} at constant pressure $P=0$~bar and temperature $T=1342$~K (NPT ensemble), which corresponds to an undercooling of $\frac{\Delta T}{T_m}\approx20\%$. The supercooled liquid was obtained by melting the $\text{Ni}_3\text{Al}$ fcc bulk supercell at $T=2500$~K and a subsequent, continuous cooling to the target undercooling of 20\%. For every structure, 1000 snapshots taken every 500~ps were included from the corresponding MD trajectories to account for the effect of thermal fluctuations in the map. Every point in the $\bar{q}_4-\bar{q}_6$ map shown in Figure~\ref{fig:ni3alq4q6} corresponds to an atom in the snapshots for a particular structure.
%
\section{Structural Composition of Presumed Transition State Clusters}
\begin{figure}
\centering
\begin{subfigure}{0.45\linewidth}
\caption{}
\label{fig:subns320Q639str}
  \includegraphics[width=0.9\linewidth]{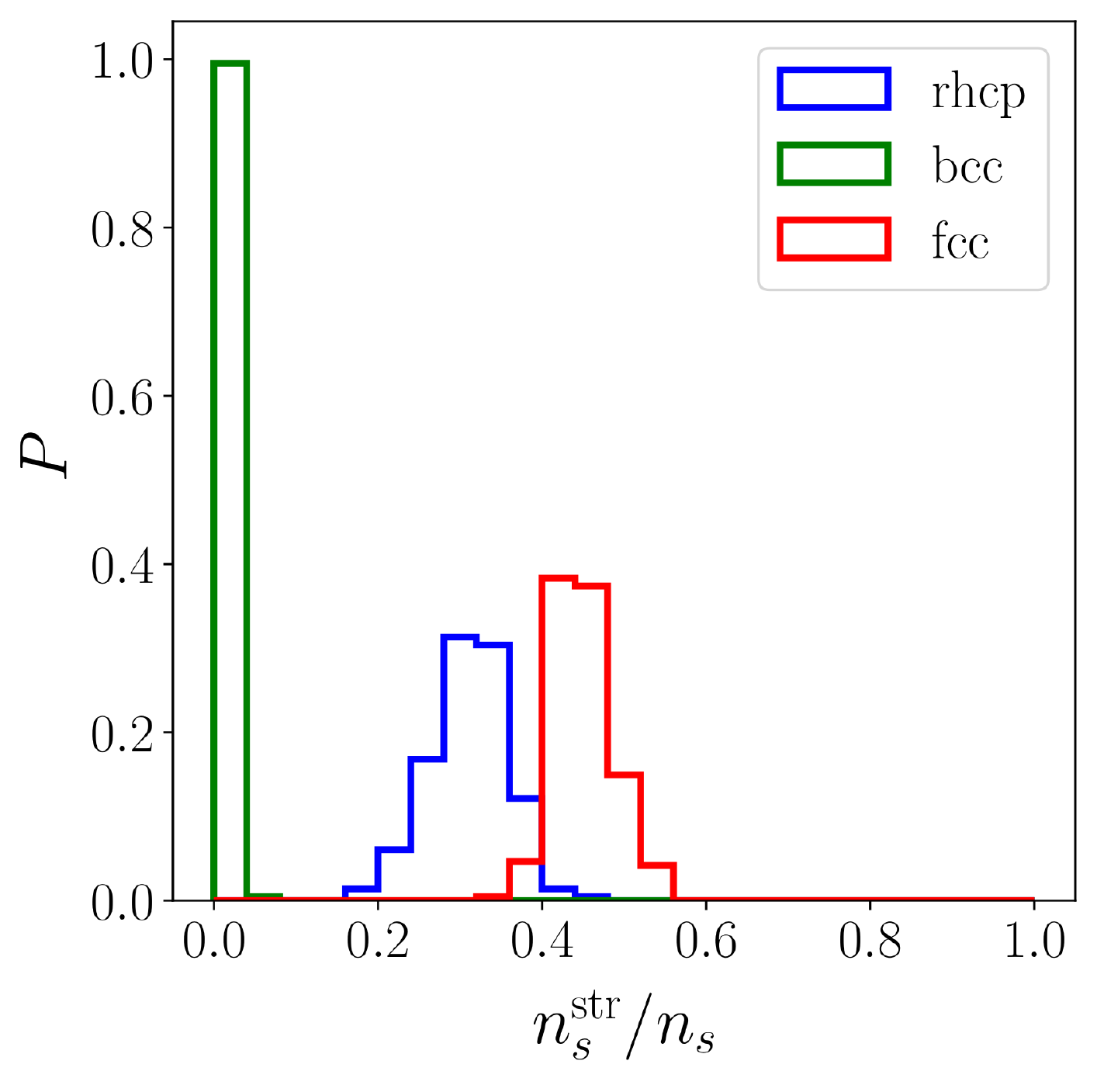}
\end{subfigure}
\hfill
\begin{subfigure}{0.45\linewidth}
\caption{}
\label{fig:subns458Q633str}
  \includegraphics[width=0.9\linewidth]{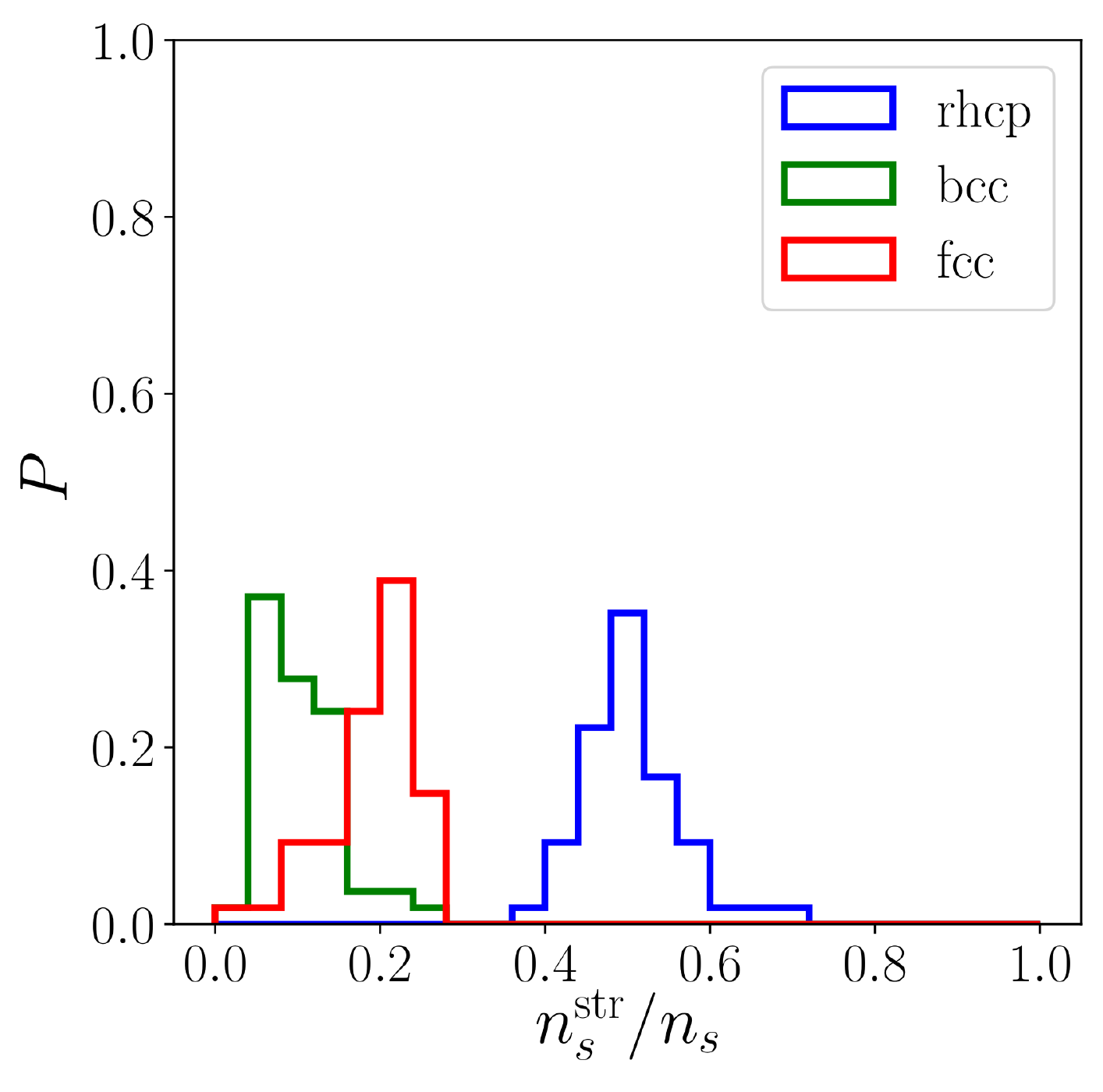}
\end{subfigure}
\caption{Structural composition of two types of TS nuclei with $\overline{p}_{B}(n_{s},Q_{6}^{cl})\approx0.5$. Fcc, bcc, and rhcp are colored in red, green, and blue, respectively. (a) TS nuclei with $n_{s}^*=320,Q_{6}^{cl*}\geq0.39$. These clusters are predominantly fcc with some rhcp and little bcc.
(b) TS nuclei with $n_{s}^*=458,Q_{6}^{cl*}\leq0.33$. These clusters are composed of a mixture of fcc, bcc, and rhcp. The amount of fcc and bcc is comparable.}
\label{fig:TSlargesmallnsstr}
\end{figure}

The structural composition for two types of presumed critical clusters with an averaged committor  $\overline{p}_{B}(n_{s},Q_{6}^{cl})\approx0.5$ is analyzed. The corresponding distribution for 724 small/high-crystallinity clusters ($n_{s}^*=320,Q_{6}^\text{cl*}\geq0.39$) obtained from the reweigthed path ensemble (RPE) is shown in Figure~\ref{fig:subns320Q639str}. The small/high-crystallinity clusters are composed of mainly fcc with negligible bcc, and some rhcp. In comparison, the distribution for 137 large/low-crystallinity clusters ($n_{s}^*=458,Q_{6}^\text{cl*}\leq0.33$) in Figure~\ref{fig:subns458Q633str} shows a mixture of rhcp, fcc and bcc with approximately 50\% rhcp and a comparable amount of fcc and bcc with $10-20$\%. 
%
\section{Committor Distribution and Chemical Order of Large/Low-Crystallinity Clusters}

\begin{figure}
\centering
\begin{subfigure}{0.43\linewidth}
\caption{}
\label{fig:ns458pbhisto}
 \includegraphics[width=\textwidth]{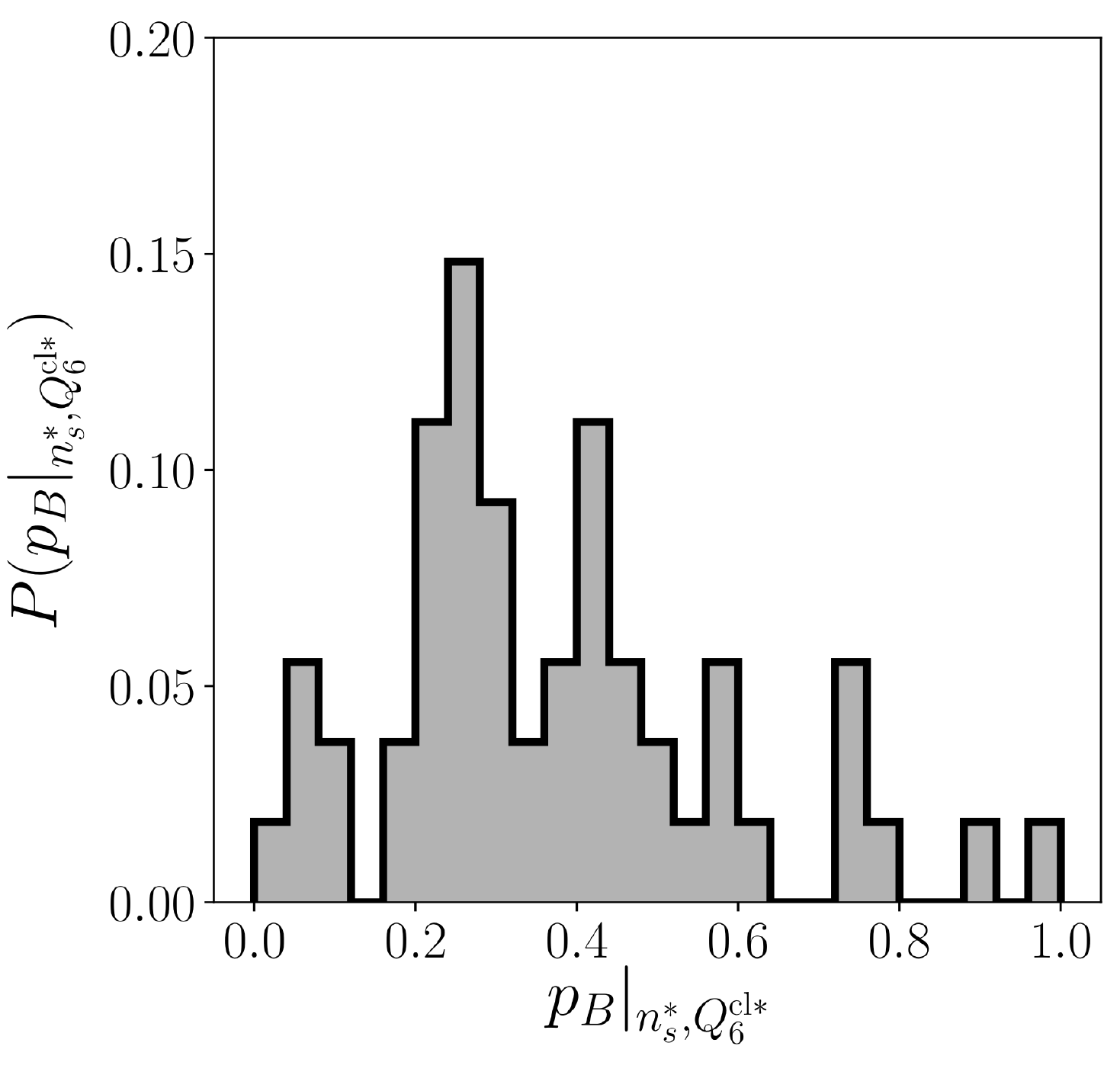}
 \end{subfigure}
\hfill
\begin{subfigure}{0.49\linewidth}
\caption{}
\label{fig:ns458sro1sro2pb}
  \includegraphics[width=\linewidth]{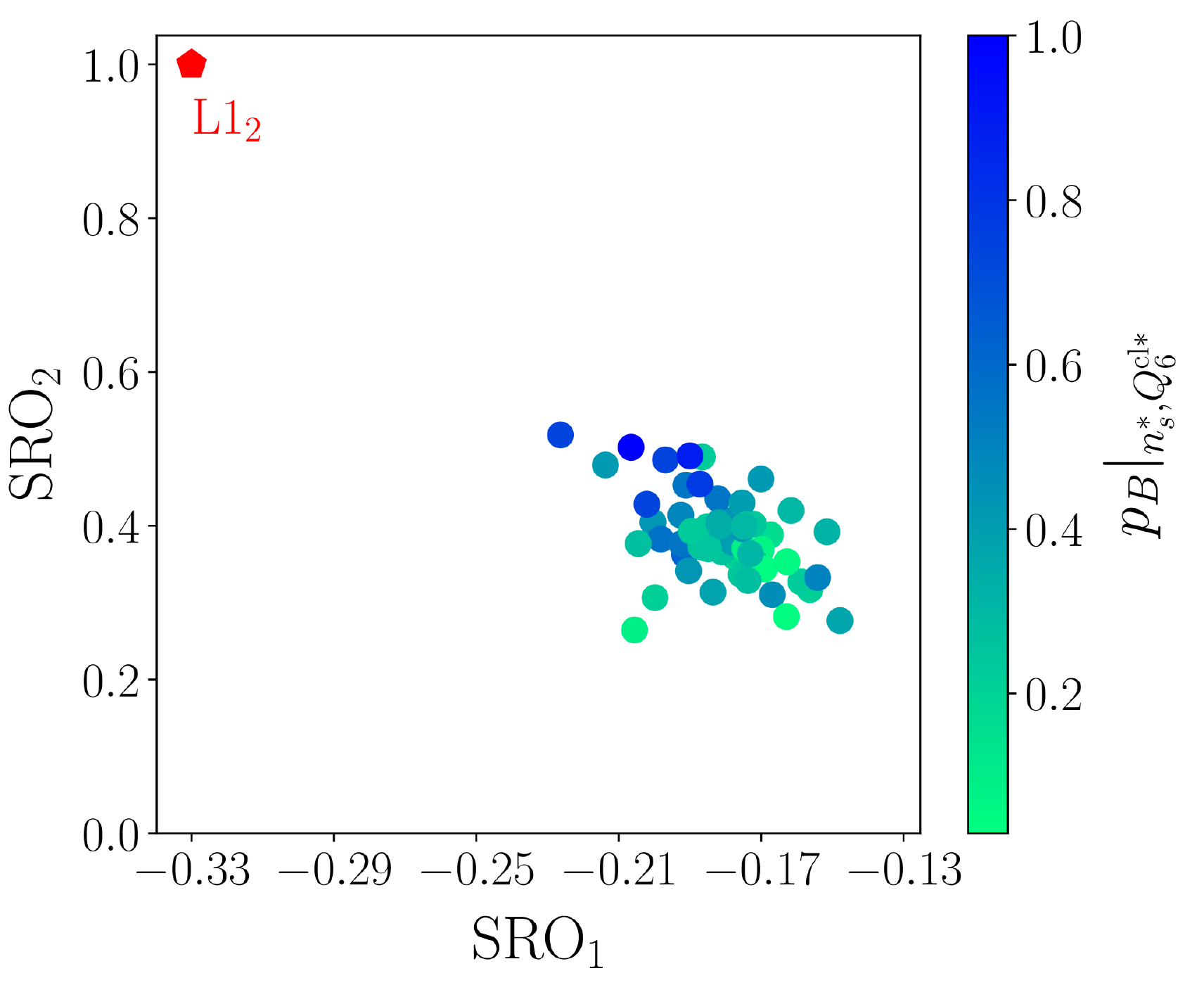}
\end{subfigure}
\caption{ (a) $P(p_B|_{ n_s^*, Q_6^\text{cl*}})$ computed from 54 (assumed) TS large/low-crystallinity clusters satisfying $n_{s}^*=458$, $Q_{6}^\text{cl*} \leq 0.33$. (b) Committor values of assumed TS configurations for large/low-crystallinity clusters with $n_{s}^* = 458$ and $Q_{6}^\text{cl*} \leq 0.33$ as a function of the chemical short-range order. The red pentagon marks the reference parameters, $\text{SRO}_1$ and $\text{SRO}_2$, of $\text{L1}_2$.
}
\end{figure}

The committor distributions of large/low-crystallinity clusters were obtained from 54 configurations satisfying $n_{s}^*=458$, and $Q_{6}^\text{cl*} \leq 0.33$. In Figure~\ref{fig:ns458pbhisto}, the committor distribution shows a spread over the whole range, with an increase in the region $0.2<{p}_{B}<0.6$. The correlation of chemical short-range order, $\text{SRO}_1$ and $\text{SRO}_2$, with the corresponding committor values ${p}_{B}$ are shown in Figure~\ref{fig:ns458sro1sro2pb}. Since these large/low-crystallinity clusters are mostly rhcp and do not have well defined, dominant crystalline structures, they exhibit mostly moderate chemical SROs in the first and second nearest neighbour shells and only very little correlation of SRO with $p_B$ is observed.
Considering the presence of bcc and rhcp, the SROs are further evaluated specifically within bcc and hcp dominant cluster cores taking the corresponding $\text{DO}_3$ and $\text{DO}_{19}$, that are reported as metastable bcc and hcp phases in $\text{Ni}_3\text{Al}$, as references. Interestingly, the bcc and rhcp particles do not show $\text{SRO}_2$ values towards $\text{DO}_3$ and $\text{DO}_{19}$, but in the first neighboring shell Al atoms are surrounded by Ni resulting in $\text{SRO}_1$ values that are typical of the ordered phases. The SROs of bcc and hcp are barely correlated with the committor values, while SROs in fcc do correlate with the committor and increase towards $\text{L1}_2$. 
%

\section{Correlation of Collective Variables and Committor in Small, Compact Clusters}

\begin{figure}[htpb]
\centering
\includegraphics[width=1.0\textwidth]{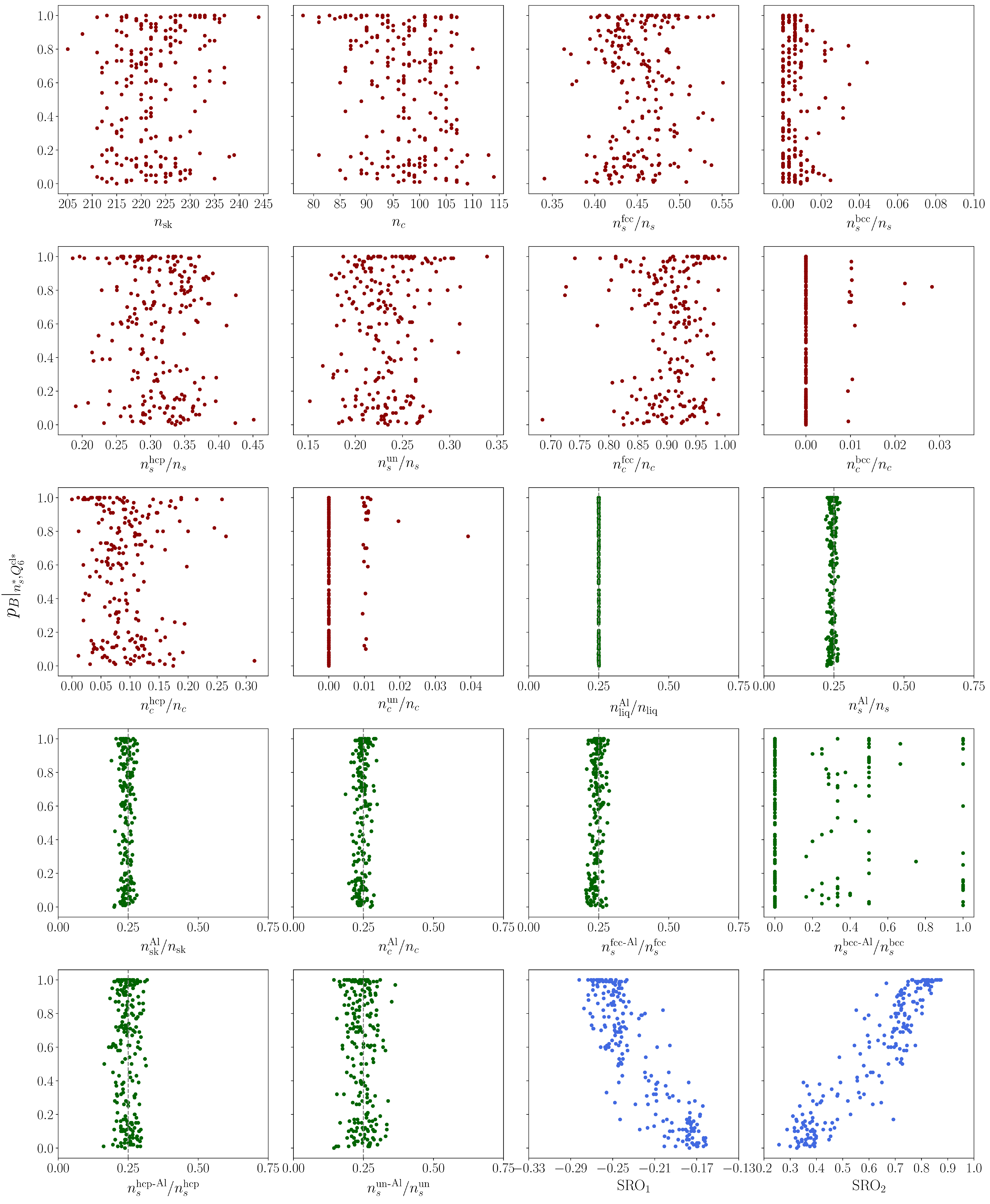}
\caption{The correlation between various CVs and the committor ${p}_{B}$ for small/high-crystallinity clusters. The structural composition (brown), chemical composition (green), and chemical SRO (blue) are presented from top left to bottom right. Every plot includes 214 points.}
\label{fig:allcorrelation}
\end{figure}

To check if there is an additional collective variable (CV) that improves the description of the RC for the crystal nucleation in $\text{Ni}_3\text{Al}$, the correlation between the committor probability and relevant CVs is closely examined for small, compact clusters.
20 CVs concerning the structural composition, chemical composition, and chemical SROs are analyzed for small/high-crystallinity clusters (214 configurations) satisfying $n_{s}^* = 320\pm5$, and $Q_{6}^\text{cl*} \geq 0.39$. The committor values as a function of these CVs are plotted in Figure~\ref{fig:allcorrelation}. All CVs related to the structural composition of the clusters (see section~\Romannum{2}~C~1 and section~\Romannum{2}~C~2 in the main text) show no additional correlation with the committor (graphs with brown symbols in Figure~\ref{fig:allcorrelation}). 
The global crystallinity $Q_{6}^\text{cl}$ appears to  already capture the important structural features.
The chemical composition within various parts of the clusters does not improve the description of RC as well, showing little correlation with the committor (graphs with green symbols in Figure~\ref{fig:allcorrelation}).
The amount of Al is always close to 25\% with committor values over the entire range from 0 to 1. 
Since there are almost no bcc particles in these small compact clusters, the resulting Al composition in the bcc phase, $\frac{n_s^\text{bcc-Al}}{n_s}$, is mostly zero.
The CVs describing the chemical short-range order in the clusters exhibit, however, a strong correlation with the committor (graphs with blue symbols in Figure~\ref{fig:allcorrelation}). Specifically, $p_{B}$ increases with increasing values of $\text{SRO}_1$ and $\text{SRO}_2$ towards $\text{L1}_2$ ordering ($\text{SRO}_1=-0.33$, and $\text{SRO}_2=1$).
The strong correlation of the chemical short-range order with the committor indicates that the SROs of the cluster should be considered as a possible addition to the RC model in $\text{Ni}_3\text{Al}$ combined with the size and crystallinity.
%

\section{Solidification Simulations}

\begin{figure}[htpb]
\centering
\begin{subfigure}{0.45\linewidth}
\caption{}
\label{fig:slequil}
 \includegraphics[width=\textwidth]{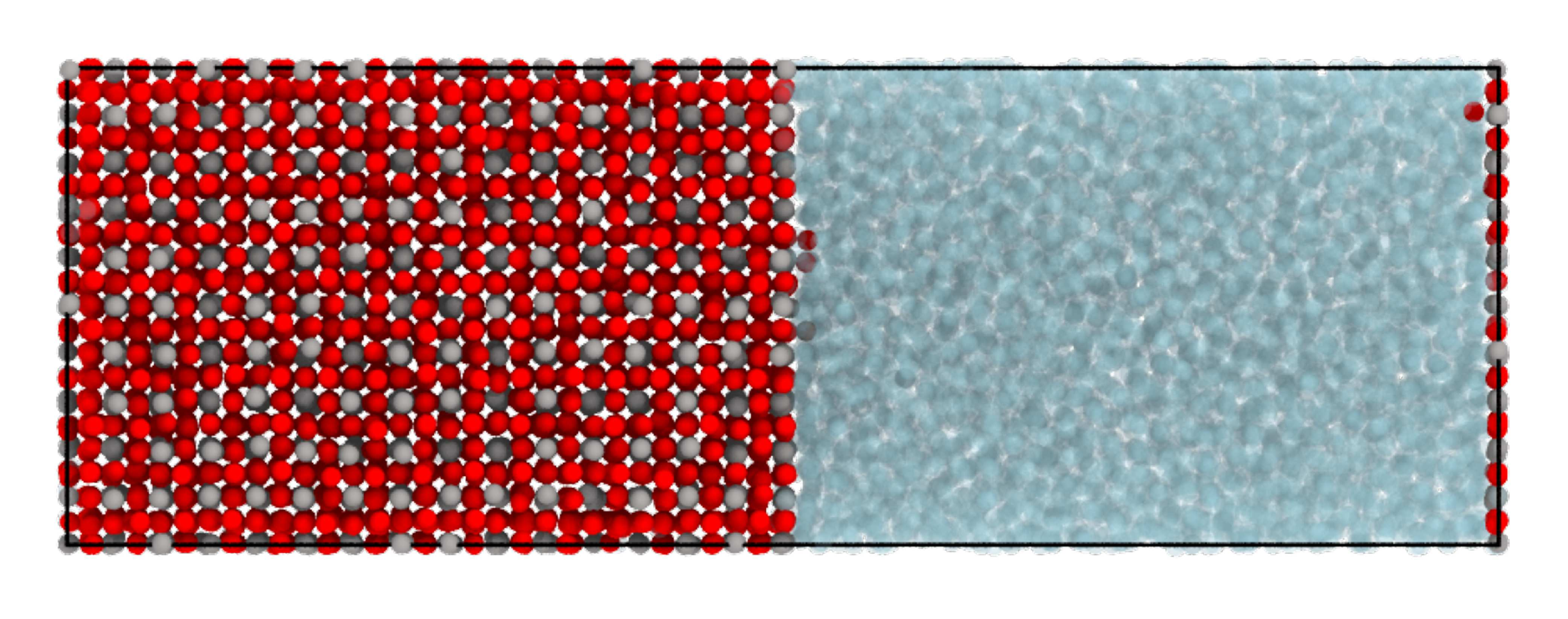}
 \end{subfigure}
\hfill
\begin{subfigure}{0.45\linewidth}
\caption{}
\label{fig:sl1ns}
  \includegraphics[width=\linewidth]{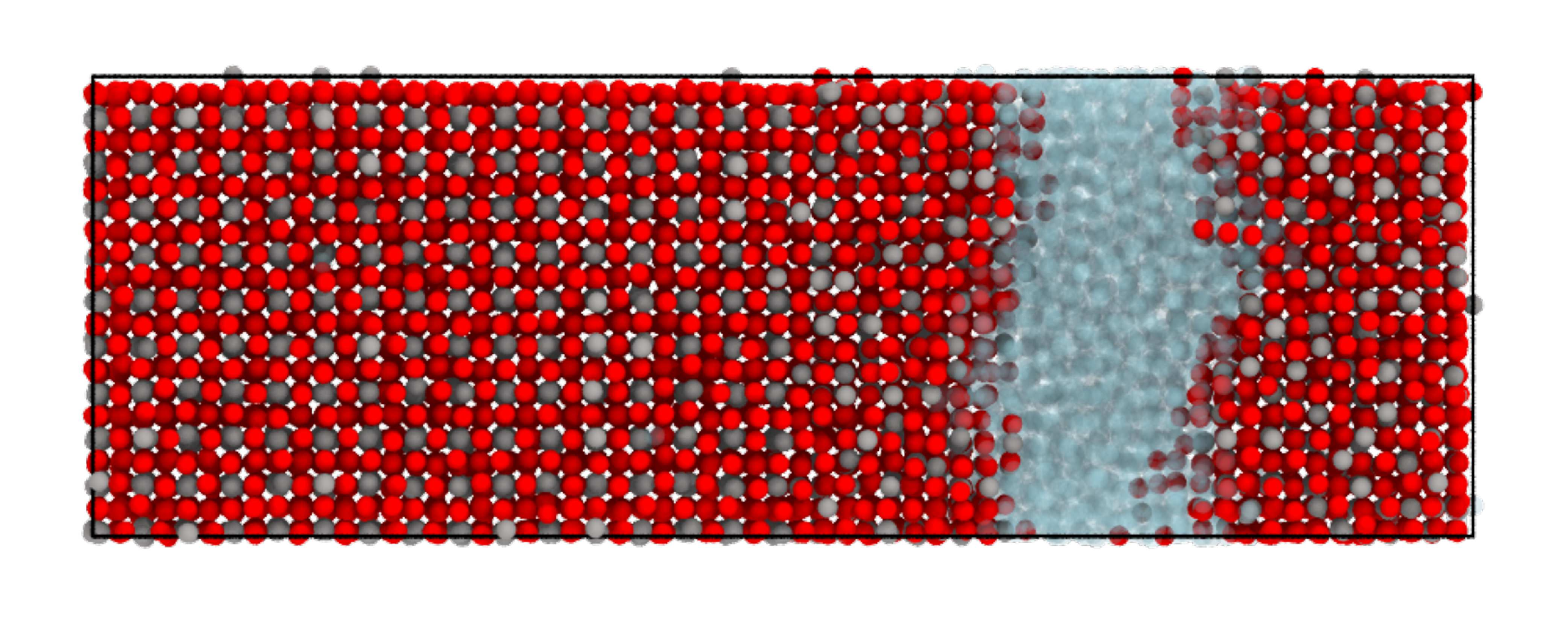}
\end{subfigure}
\caption{The solid-liquid equilibrium system composed of $\text{L1}_2$ ordered bulk phase (Ni in red, Al in gray) and liquid $\text{Ni}_3\text{Al}$ (transparent blue). (a) Snapshot of the equilibrated solid-liquid interface at 1\% undercooling at $\delta t=0$. (b) Snapshot after $\delta t=1$~ns at constant undercooling of 1\%.}
\label{fig:solidliquid}
\end{figure}

\begin{figure}
\centering
\begin{subfigure}{0.45\linewidth}
\caption{}
\label{fig:liquidnsgrow}
 \includegraphics[width=0.9\textwidth]{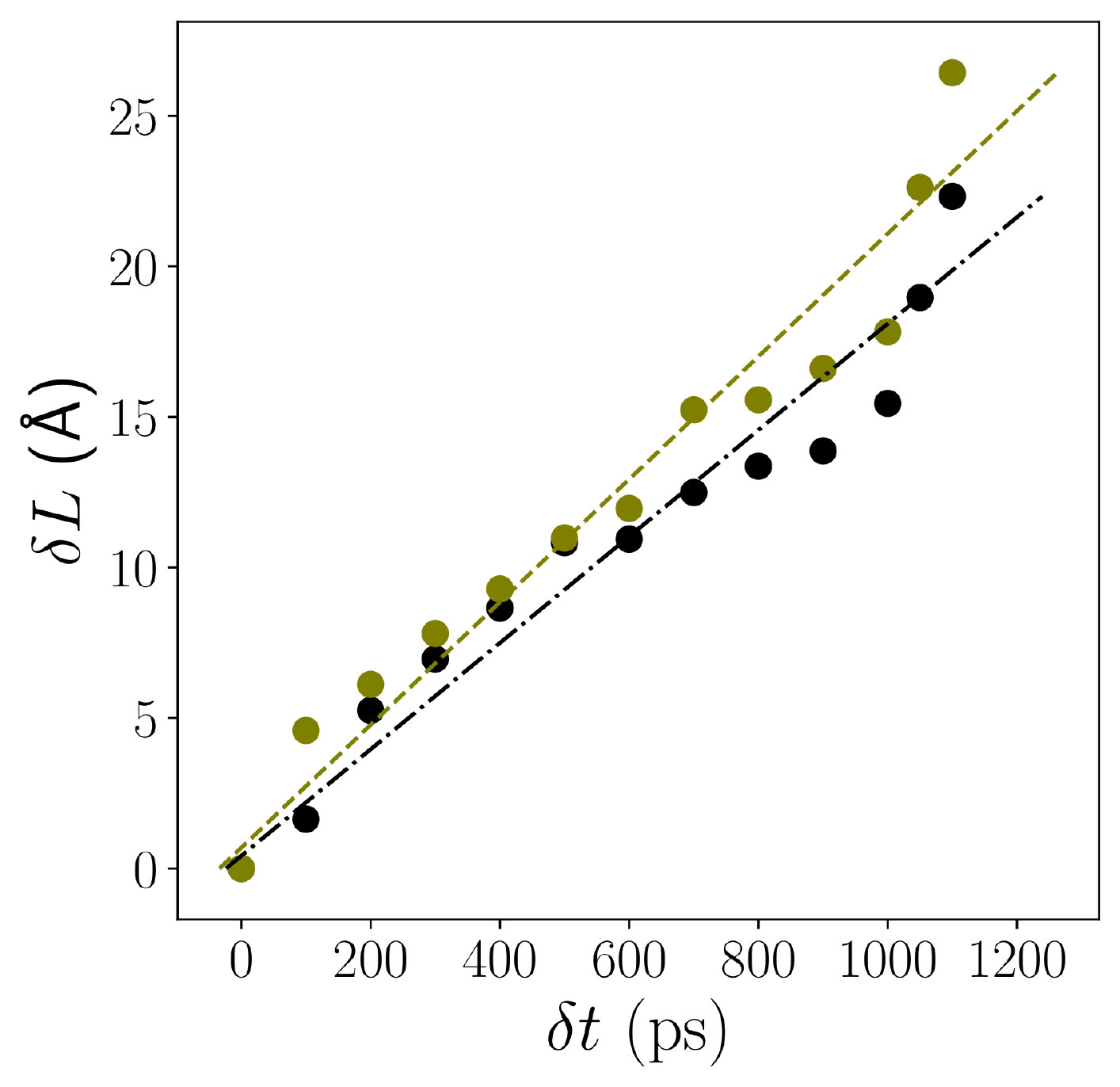}
 \end{subfigure}
\hfill
\begin{subfigure}{0.45\linewidth}
\caption{}
\label{fig:sroliquid}
  \includegraphics[width=0.9\linewidth]{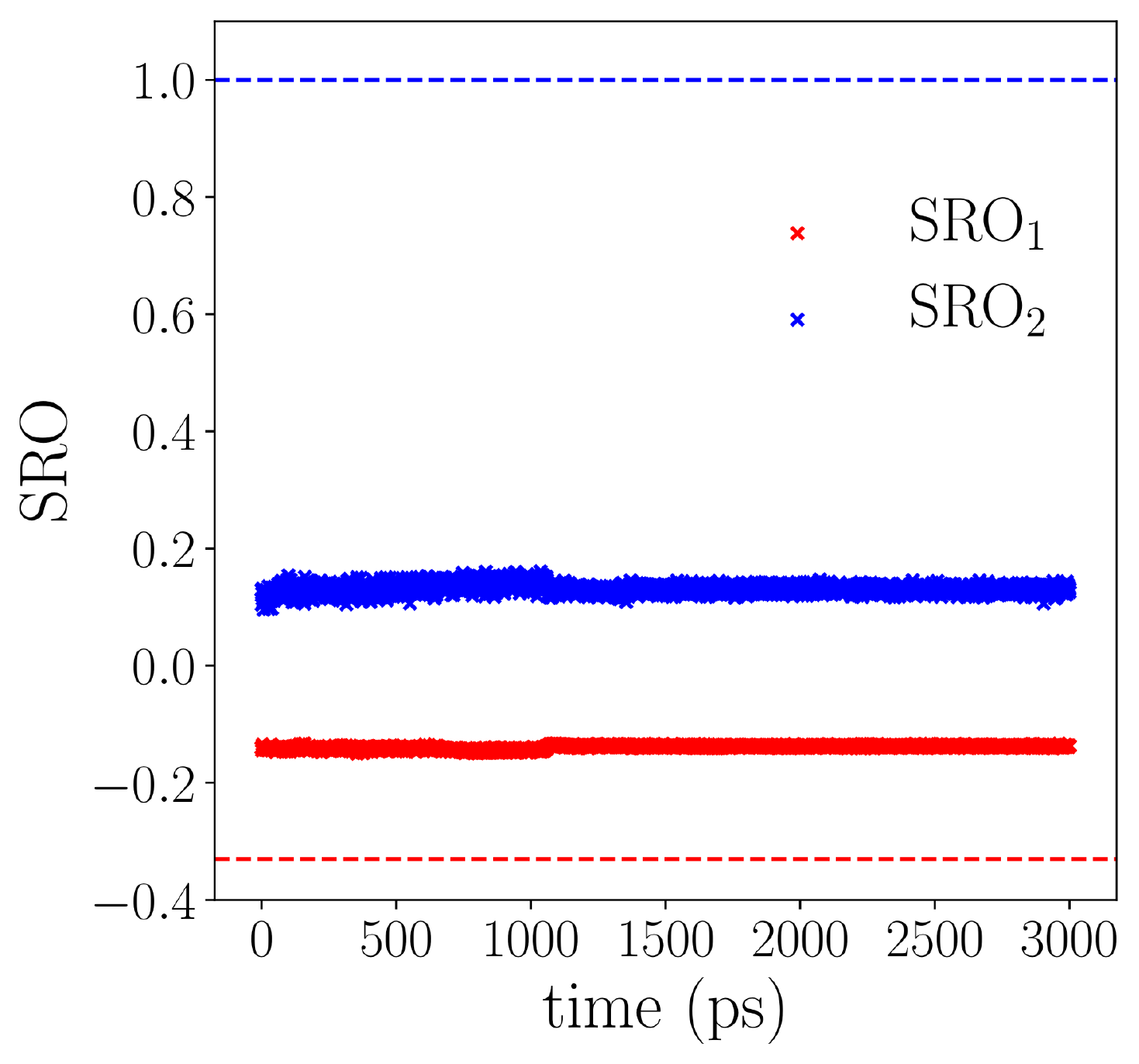}
\end{subfigure}
\caption{ (a) Change of interface positions as a function of time. Dashed lines are linear fits to the data. (b) Evolution of the chemical short-range order in the solidified bulk region as a function of time. The dashed lines represent the reference values for $\text{SRO}_1$ (red dashed line) and $\text{SRO}_2$ (blue dashed line) in $\text{L1}_2$. }
\end{figure}

To test the impact of chemical SRO during the growth stage in $\text{Ni}_3\text{Al}$, we conducted MD simulations of solidification initiated from an equilibrated interface between $\text{L1}_2$ ordered bulk and liquid.
The initial simulation box was constructed with $\text{L1}_2$ bulk phase containing 12000 atoms (9000 Ni, and 3000 Al atoms) corresponding to a $15\times10\times10$ supercell. The whole system was equilibrated initially at $P=0$~bar, $T=1660$~K within the NPT ensemble, 18~K below the melting temperature, for 200~ps. A subsequent melting was applied to half of the simulation box at $T=3300$~K for 2~ns while the atoms in the other half were fixed. Hereafter, the whole system was brought to the melting temperature of $\text{Ni}_3\text{Al}$ at $T=1678$~K and equilibrated for another 2~ns. In this way, an equilibrated system having an $\text{L1}_2$-liquid interface in [001] direction was created.  
After equilibration, the system was kept at $T=1660$~K, approximately 1\% undercooling, for another 3~ns and continuous solidification was observed. Figure~\ref{fig:solidliquid} illustrates the configurations obtained at different stages: a snapshot of the system before solidification at $\delta t=0$, and a snapshot of the growth of solid bulk at $\delta t=1$~ns.
During the growth stage, the position of the solid-liquid interface was closely monitored and the interface velocity was calculated as $v=\frac{\delta L}{\delta t}$ from an estimation of the change of interface position $\delta L$ averaged over time $\delta t$. The average interface velocity was approximated to be $\bar{v}\approx3\pm 1.2$~m/s from 5 MD simulations. 
In Figure~\ref{fig:liquidnsgrow}, the change of the two interface positions as a function of time extracted from one MD simulation is shown. Along the solidification, the chemical SROs were also analyzed within the solidified part.
Figure~\ref{fig:sroliquid} shows that the short-range order, both SRO$_1$ and SRO$_2$, in the solidified part is similar to the one in the melt and does not correspond to an $\text{L1}_2$ ordering. The bulk thus inherits the chemical order of the melt as a result of the large interface velocity. The fast growth from the interface surpasses the mobility of the atoms in the melt, and thus the atoms do not have enough time to rearrange into the thermodynamically favoured $\text{L1}_2$ phase, that is we observe disorder trapping during the growth stage.~\cite{boettinger1989theory,aziz1994transition}
%

\bibliography{NiRC}